\documentclass[10pt,a4paper,reqno]{amsart}
\usepackage{acronym}
\usepackage{amsfonts}
\usepackage{amsmath}
\usepackage{amssymb}
\usepackage{amsthm}
\usepackage{bm}
\usepackage{braket}
\usepackage{cite}
\usepackage{color}
\usepackage{geometry}
\usepackage{graphicx}
\usepackage{hyperref}
\usepackage{mathrsfs}
\usepackage{mathtools}
\usepackage{setspace}

\usepackage{booktabs}
\usepackage{threeparttable}
\usepackage{diagbox}
\usepackage{multirow}
\usepackage{float}
\usepackage{array}
\usepackage{graphicx}

\usepackage{stmaryrd}
\usepackage{subcaption}
\usepackage{tikz}

\usepackage{epsfig}
\usepackage{setspace}
\usepackage{epstopdf}
\usepackage{booktabs}
\usepackage{threeparttable}
\usepackage{diagbox}

\usepackage{bbm, dsfont}

\newgeometry{top=2cm,bottom=2cm,outer=1.5cm,inner=1.5cm}
\newcommand{\bse}{\begin{subequations}}
\newcommand{\ese}{\end{subequations}}

\newtheorem{theorem}{Theorem}

\newtheorem{remark}[theorem]{Remark}

\numberwithin{equation}{section}
\DeclareSymbolFont{largesymbols}{OMX}{yhex}{m}{n}
\DeclareMathAccent{\Widehat}{\mathord}{largesymbols}{"62}

\usepackage[linesnumbered,ruled,vlined]{algorithm2e}
\usepackage{amsmath}

\SetNlSty{textbf}{}{}  
\SetAlgoNlRelativeSize{0}  
\SetNlSkip{2em}  
\SetAlgoNlRelativeSize{-1}

\title[Causality-guided adaptive sampling method for physics-informed neural networks]{Causality-guided adaptive sampling method for physics-informed neural networks}
\author{Shuning Lin}
\address[SL]{School of Mathematical Sciences, Shanghai Key Laboratory of Pure Mathematics and Mathematical Practice, and Shanghai Key Laboratory of Trustworthy Computing \\
East China Normal University \\ Shanghai 200241 \\ China}
\author{Yong Chen$^*$}
\address[YC]{School of Mathematical Sciences, Shanghai Key Laboratory of Pure Mathematics and Mathematical Practice, and Shanghai Key Laboratory of Trustworthy Computing \\
East China Normal University \\ Shanghai 200241 \\ China}
\address[YC]{College of Mathematics and Systems Science \\ Shandong University of Science and Technology \\ Qingdao 266590 \\ China}
\email{ychen@sei.ecnu.edu.cn}

\begin{document}

\begin{abstract}

Compared to purely data-driven methods, a key feature of physics-informed neural networks (PINNs) — a proven powerful tool for solving partial differential equations (PDEs) — is the embedding of PDE constraints into the loss function. The selection and distribution of collocation points for evaluating PDE residuals are critical to the performance of PINNs. Furthermore, the causal training is currently a popular training mode. In this work, we propose the causality-guided adaptive sampling (Causal AS) method for PINNs. Given the characteristics of causal training,  we use the weighted PDE residuals as the indicator for the selection of collocation points to focus on areas with larger PDE residuals within the regions being trained. For the hyper-parameter $p$ involved, we develop the temporal alignment driven update (TADU) scheme for its dynamic update beyond simply fixing it as a constant. The collocation points selected at each time will be released before the next adaptive sampling step to avoid the cumulative effects caused by previously chosen collocation points and reduce computational costs. To  illustrate the effectiveness of the Causal AS method, we apply it to solve time-dependent equations, including the Allen-Cahn equation, the NLS equation, the KdV equation and the mKdV equation. During the training process, we employe a time-marching technique and strictly impose the periodic boundary conditions by embedding the input coordinates into Fourier expansion to mitigate optimization challenges. Numerical results indicate that the predicted solution achieves an excellent agreement with the ground truth. Compared to a similar work, the causal extension of R3 sampling (Causal R3), our proposed Causal AS method demonstrates a significant advantage in accuracy.


\end{abstract}
\maketitle

\section{Introduction}

In recent years, the physics-informed neural network (PINN) \cite{PINN} proposed for solving forward and inverse problems of partial differential equations (PDEs) has rapidly developed and has become one of the most promising frameworks. It seamlessly integrates data with physics models, which enhances the accuracy and reliability of solutions. As a mesh-free approach, the PINN facilitates the resolution of high-dimensional problems and overcomes the curse of dimensionality. This mesh-free nature also allows for greater adaptability to complex geometries and boundary conditions without the necessity of custom-designed grids for each problem, thus significantly improving computational efficiency. The PINN algorithm and its extensions can be applied to different types of equations, including integro-differential equations \cite{DeepXDE-RAR}, fractional-order systems \cite{fPINN}, stochastic differential equations (SDE)\cite{SDE1, SDE2} and discretization equations \cite{discretesystems}. Despite achieving significant success, PINNs still encounter challenges when dealing with certain complex problems, leading to the emergence of numerous variants tailored to different requirements. Currently, there are five main aspects of improvement for PINN. (1) Modifying the network architecture: Wang et al. put forth the modified MLP architecture \cite{modifiedMLP} and Fourier feature networks \cite{WSFNTKFourier} designed to address high-frequency and multi-scale problems, Cheng et al. proposed Res-PINN \cite{Res-PINN} combines the PINN with Resnet blocks; (2) Domain decomposition: There are many effective methods that decompose the temporal domain \cite{AdaptivePINN, bcPINN}, segment the spatial domain \cite{cPINN}, and address both temporal and spatial domains simultaneously \cite{XPINN}; (3) Loss modification: More physical constraints can be embedded into the loss function, such as conserved quantities \cite{LSNJCP}, symmetries \cite{PINNsymmetries} and Lax pair \cite{LaxPairPINN}; (4) Loss balancing: Adaptive weights \cite{modifiedMLP,SA-PINNs,weight-NTK} are designed to balance the effects of different loss terms; (5) Sampling of residual collocation points: Residual-based adaptive sampling is the most widely adopted strategy \cite{DeepXDE-RAR,RADandRAR-D}. This paper mainly focuses on the final aspect.

The primary distinction between the PINN and purely data-driven neural network methods lies in the construction of the loss function: PINN incorporates PDE constraints, requiring the selection of a certain number of collocation points for computing the PDE residuals. The selection of residual points is analogous to the grid cells in the finite element method, suggesting that the distribution of collocation points plays a crucial role in PINN. Moreover, an excessive number of residual points can hinder optimization. Therefore, identifying the optimal or more favorable distribution of collocation points and establishing the principles for selection have consistently been major research focuses. Initially, Latin hypercube sampling method \cite{LHS}, a type of stratified sampling, was used to select collocation points, achieving approximate random sampling from a multivariate parameter distribution. Subsequently, adaptive sampling strategies are developed, which can be categorized into the following types. Residual-based adaptive sampling scheme is mainly based on the idea of using the residual (referring to the PDE loss) to indicate the error (the difference between the predicted solution and the true solution), given that the residual is a directly accessible quantity. Representative methods include the residual-based adaptive refinement (RAR) \cite{DeepXDE-RAR}, residual-based adaptive distribution (RAD) \cite{RADandRAR-D} and residual-based adaptive refinement with distribution (RAR-D) \cite{RADandRAR-D}. Additionally, the FI-PINN method \cite{FIPINN} defines an effective failure probability based on the residuals and employs a failure-informed enrichment technique to adaptively add new collocation points. Gradient-based adaptive sampling scheme is designed to help to relocate the collocation points to areas with higher loss gradient. Ref. \cite{gradient-based-sample-1} gradually changed the sampling of collocation points from uniform to adaptive through a cosine annealing strategy. To enhance the accuracy of solving problems with sharp solutions, Liu et al. \cite{gradient-based-sample-2} proposed the EI-Grad algorithm, which is built upon the EI-RAR algorithm by incorporating gradient information of the residual values as the criterion for selecting sampling points. Related works also include \cite{gradient-based-sample-3,gradient-based-sample-4} and others. Another approach considers the limitations of using residuals as error indicators, particularly the need for derivative calculations, which can be highly expensive computationally. Chen et al. put forward the adaptive trajectories sampling (ATS) technique \cite{trajectories-based-sample}, where the training points are selected adaptively according to an empirical-value-type indicator from trajectories  generated by a PDE-related stochastic process.

Research on PINNs has increasingly highlighted the importance of respecting temporal causality. The evolution of physical systems adheres to the principle of causality, whereby the present state is dictated by its preceding conditions. However, the existing residual loss formula of PINNs simultaneously optimizes the residuals of partial differential equations at all time points. Given that the inability of existing PINNs formulations to respect the temporal causal structure — a key source of error leading to convergence to erroneous solutions — the causal training for PINN (Causal PINN) \cite{CausalPINN} and the causality-enforced evolutional network (CEEN) \cite{CEEN} were designed to address this issue, resulting in significant accuracy improvements. The principle behind these methods is to ensure that the model is well-trained for earlier time points before progressively addressing later ones when training PINN. In this study, we aim to propose an adaptive sampling method respecting causality that leverages the advantages of temporal causality in the design of collocation point selection. Our main contributions can be summarized as follows.

\begin{itemize}
 \item  In the Causal PINN framework, given the presence of a temporal weight function designed to explicitly adhere to the causal structure, we propose the causality-guided adaptive sampling (Causal AS) method, which is a novel method using weighted residuals as the criterion for selecting collocation points. Another key feature of our method is that newly selected collocation points are released before the next adaptive sampling step. It can avoid the cumulative bias of previously chosen points and effectively reduces the training cost.

 \end{itemize}

\begin{itemize}
 \item In the design of the weight function, we introduce a hyper-parameter $p$ that governs the distribution of sampling points by modulating the dominance between the weights and the PDE residuals. Beyond the straightforward approach of fixing $p$ as a constant, we also design the temporal alignment driven update (TADU) scheme for dynamically updating $p$.

 \end{itemize}

\begin{itemize}
 \item We extend the original Causal PINN framework to be suitable for solving problems with non-periodic boundary conditions by re-formulating the loss function and adjusting the temporal weights. 
 
 \end{itemize}
 
\begin{itemize}
 \item The effectiveness of our proposed strategy is demonstrated through extensive numerical experiments on solving time-dependent partial differential equations. Furthermore, we compare the performance of this method with a similar work, causal extension of R3 sampling (Causal R3) \cite{CausalR3} to highlight the significant advantage in accuracy of the Causal AS method.  
\end{itemize}

This paper is organized as follows. In Section \ref{Methodology}, we put forward the causality-guided adaptive sampling method tailored for the causal training. In Section \ref{experiments}, we apply the Causal AS method to address the forward problems of four partial differential equations: the Allen-Cahn equation, the NLS equation, the KdV equation, and the mKdV equation, while also evaluating its effectiveness in comparison to the Causal R3 method. Finally, Section \ref{Conclusion} concludes the paper and offers perspectives for future research.

\section{Methodology}\label{Methodology}
\subsection{Causal physics-informed neural networks}
\quad

Let's first briefly review the framework of solving the forward problem using physics-informed neural networks (PINNs) \cite{PINN}. Consider the following general form of the evolution equation,
\begin{align}
\boldsymbol{u}_t+\mathcal{N}[\boldsymbol{u}]=0, \quad t \in[T_0, T_1], \quad \boldsymbol{x} \in \boldsymbol{\Omega},
\end{align}
given the initial and boundary conditions as
\begin{align}
&\boldsymbol{u}(T_0, \boldsymbol{x})=\boldsymbol{g}(\boldsymbol{x}), \quad \boldsymbol{x} \in \boldsymbol{\Omega}, \\
&\mathcal{B}[\boldsymbol{u}]=0, \quad t \in[T_0, T_1], \quad \boldsymbol{x} \in \boldsymbol{\partial \Omega}.
\end{align}
Here, $\mathcal{N}[\cdot]$ denotes the differential operator with respect to the spatial variables and $\mathcal{B}[\cdot]$ is the boundary operator. Then a deep neural network $\boldsymbol{u}_{\boldsymbol{\theta}}(t, \boldsymbol{x})$, for instance, with a depth of $L$, is trained to approximate the solution $\boldsymbol{u}(t, \boldsymbol{x})$ of the PDE, where $\boldsymbol{\theta}=\left\{\mathbf{w}^{l}, \mathbf{b}^{l}\right\}_{l=1}^{L}$ represent the trainable parameters. The default architecture of the neural network is multi-layer perceptrons (MLPs), also known as fully-connected feedforward neural networks, since their expressive power guaranteed by the universal approximation theorem\cite{MLP}. Based on the given physics and data information, we randomly sample the sets of initial-boundary points and residual points ($\left\{\boldsymbol{x}_{i c}^i\right\}_{i=1}^{N_{i c}},\left\{t_{b c}^i, \boldsymbol{x}_{b c}^i\right\}_{i=1}^{N_{b c}}$ and $\left\{t_r^i, \boldsymbol{x}_r^i\right\}_{i=1}^{N_r}$), and proceed by training the physics-informed neural network through minimizing the mean squared error loss
\begin{align}
\mathcal{L}(\boldsymbol{\theta})=\lambda_{i c} \mathcal{L}_{i c}(\boldsymbol{\theta})+\lambda_{b c} \mathcal{L}_{b c}(\boldsymbol{\theta})+\lambda_r \mathcal{L}_r(\boldsymbol{\theta})
\end{align}
where
\begin{align}
&\mathcal{L}_{i c}(\boldsymbol{\theta})=\frac{1}{N_{i c}} \sum_{i=1}^{N_{i c}}\left|\boldsymbol{u}_{\boldsymbol{\theta}}\left(T_0, \boldsymbol{x}_{i c}^i\right)-\boldsymbol{g}\left(\boldsymbol{x}_{i c}^i\right)\right|^2,\\
&\mathcal{L}_{b c}(\boldsymbol{\theta})=\frac{1}{N_{b c}} \sum_{i=1}^{N_{b c}}\left|\mathcal{B}\left[\boldsymbol{u}_{\boldsymbol{\theta}}\right]\left(t_{b c}^i, \boldsymbol{x}_{b c}^i\right)\right|^2,\\
&\mathcal{L}_r(\boldsymbol{\theta})=\frac{1}{N_r} \sum_{i=1}^{N_r}\left|\frac{\partial \boldsymbol{u}_{\boldsymbol{\theta}}}{\partial t}\left(t_r^i, \boldsymbol{x}_r^i\right)+\mathcal{N}\left[\boldsymbol{u}_{\boldsymbol{\theta}}\right]\left(t_r^i, \boldsymbol{x}_r^i\right)\right|^2.
\end{align}
The weight coefficients ($\lambda_{i c}, \lambda_{b c}$ and $\lambda_r$) of each loss term guide the optimization towards the overall objective, thereby achieving a balance among different goals and enhancing the stability of the training process. By selecting effective optimization algorithms, such as Adam \cite{Adam} or L-BFGS \cite{LBFGS}, the weights and biases in the neural network are iteratively updated until the algorithm converges. During the network training process, the involved partial derivatives and the gradients of the loss function with respect to the trainable parameters can be derived via automatic differentiation \cite{AD}.

Although the mesh-free PINN method has achieved significant success in solving many PDEs, especially in dealing with noisy data, complex regions and high-dimensional problems, it still faces considerable challenges when addressing multi-scale and high-frequency issues. Wang et al. \cite{CausalPINN} observed that the residuals were significantly large near the initial state but then rapidly decayed to almost zero after a certain time point when using PINN to solve the Allen-Cahn equation. It inherently tends to approximate the PDE solution even before satisfying the initial conditions, which seriously violates temporal causality. Upon analysis of the objective function, it was revealed that the existing residual loss formulation optimizes the PDE residuals at all time points simultaneously, which fails to ensure that the model is well-trained for earlier time points before progressively addressing later ones. This behavior does not respect temporal causal structure of the physical system's evolution, making the PINN model prone to converging to incorrect solutions.

Based on the above considerations, the causal training algorithm for PINNs (Causal PINNs) \cite{CausalPINN} was proposed by defining a weighted residual loss
\begin{align}
\mathcal{L}_r(\boldsymbol{\theta})=\frac{1}{N_t} \sum_{i=1}^{N_t} w_i \mathcal{L}_r\left(t_i, \boldsymbol{\theta}\right)
\end{align}
where $\{t_i \}_{i=1}^{N_t} (T_0=t_1<t_2<\cdots<t_{N_t}=T_1)$ are grid points in the temporal domain, $w_1=1$ and
\begin{align}
\mathcal{L}_r(t_i, \boldsymbol{\theta})=\frac{1}{N_x} \sum_{j=1}^{N_x}\left|\frac{\partial \boldsymbol{u}_{\boldsymbol{\theta}}}{\partial t}\left(t_i, \boldsymbol{x}_j\right)+\mathcal{N}\left[\boldsymbol{u}_{\boldsymbol{\theta}}\right]\left(t_i, \boldsymbol{x}_j\right)\right|^2,\\
w_i=\exp \left(-\epsilon \sum_{k=1}^{i-1} \mathcal{L}_r\left(t_k, \boldsymbol{\theta}\right)\right), \text { for } i=2,3, \ldots, N_t. \label{original_w_i} 
\end{align}
The point set $\{\boldsymbol{x}_j \}_{j=1}^{N_x}$ represents discrete samples of the spatial domain. The weight $w_i$ decreases over time and is inversely proportional to the magnitude of the accumulated residual loss from previous time points. Consequently,  $\mathcal{L}_r\left(t_i, \boldsymbol{\theta}\right)$ will only start to be optimized once all preceding residuals have been sufficiently minimized, allowing $w_i$ to become relatively large. The parameter $\epsilon$ regulates the ease of starting training at later time points. A larger $\epsilon$ demands a greater reduction in residuals of the previous time steps to increase the weight for subsequent one. To avoid the cumbersome adjustment of hyper-parameters, an annealing strategy is adopted by selecting a sequence of values $\{\epsilon_i\}_{i=1}^k $ that incrementally increase. As the network training progresses, all the weights tend to converge to 1. Therefore, an effective stopping criterion is proposed: training can be stopped if $\min_i w_i > \delta$ after choosing a threshold parameter $\delta \in (0,1)$, which can help reduce computational costs.  Causal PINN has been proven effective in many cases where conventional PINN model failed, including the Allen–Cahn equation, the chaotic Lorenz system, the Navier–Stokes equations and so on (see Ref. \cite{CausalPINN} for further details).

\subsection{The causality-guided adaptive sampling method}
\quad

This part presents an analysis of why some classical adaptive sampling methods are no longer applicable within the Causal PINN framework after providing a review of these strategies. Subsequently, the causality-guided adaptive sampling (Causal AS) method is put forward here.

\subsubsection{Existing adaptive sampling strategies}
\quad

Several commonly used residual-based adaptive sampling methods are listed below:

\begin{itemize}
 \item \textbf{Residual-based adaptive refinement (RAR)} \cite{DeepXDE-RAR}: A certain number of points with the largest PDE residuals are selected and added to the set of collocation points at regular iteration intervals. Repeat this procedure until the total number of iterations or the total number of residual points reaches the limit. The RAR algorithm is considered a greedy algorithm since it focuses solely on points with large residuals. Thus, it is also referred to as RAR-G.
\end{itemize}

\begin{itemize}
 \item \textbf{Residual-based adaptive distribution (RAD)} \cite{RADandRAR-D}: At regular intervals of iterations, randomly sample a set of points to serve as the new collocation point set according to the probability density function (PDF)
\begin{align}\label{RAD-pdf}
p(\mathbf{x}) \propto \frac{\varepsilon^k(\mathbf{x})}{\mathbb{E}\left[\varepsilon^k(\mathbf{x})\right]}+c,	
\end{align} 
where $\varepsilon(\mathbf{x})=|f(\mathbf{x};\hat{u}(\mathbf{x}))|,\mathbf{x}=(t,\boldsymbol{x})$ is the PDE residual, and $k\geq 0$ and $c\geq 0$ are two hyper-parameters to be configured prior to training. Repeat this procedure until the total number of iterations reaches the limit. 
\end{itemize}

\begin{itemize}
 \item \textbf{Residual-based adaptive refinement with distribution (RAR-D)} \cite{RADandRAR-D}: This method is a hybrid of RAR-G and RAD. Like RAR-G, RAR-D repeatedly adds new points to the training dataset. Similar to RAD, it samples new points based on the PDF in Eq. \eqref{RAD-pdf}.
\end{itemize}

Among these, RAD is a strategy that completely replaces the collocation points, whereas RAR and RAR-D add new points without replacing existing ones. Additionally, many other adaptive sampling strategies \cite{AdaptiveSampling1,AdaptiveSampling2,AdaptiveSampling3,AdaptiveSampling4} are special cases of RAD and RAR-D or slight modifications thereof.

When adaptively selecting collocation points, those with large PDE residuals are typically considered. However, for Causal PINN, this principle would result in newly added collocation points being concentrated in the later time regions that have not been trained yet. Moreover, whether completely replacing collocation points  or continually increasing the number of collocation points based on residuals does not benefit the current causal training. Therefore, there is an urgent need to develop an adaptive sampling strategy tailored for the Causal PINN framework.

\subsubsection{The causality-guided adaptive sampling method}\label{CausalAS}
\quad

During the training process of Causal PINN, three implicit types of regions can be identified based on the corresponding characteristics:

\begin{itemize}
 \item	Well-trained regions: $w_i$ is large while $\mathcal{L}_r\left(t_i, \boldsymbol{\theta}\right)$ is small;
\end{itemize}
\begin{itemize}
 \item Regions currently being trained: $w_i$ and $\mathcal{L}_r\left(t_i, \boldsymbol{\theta}\right)$ are both large simultaneously;
\end{itemize}
\begin{itemize}
 \item Untrained or insufficiently trained regions: $w_i$ is small while $\mathcal{L}_r\left(t_i, \boldsymbol{\theta}\right)$ is large.
\end{itemize}
Incorporating points with larger PDE residuals from the first two categories of regions, particularly the second one, into the configuration points is theoretically beneficial for the causal training of the neural network. To this end, we propose the causality-guided adaptive sampling method, which has two key points.

One is that we add $N_A$ new collocation points every $K$ iterations but these newly selected points are released before the next adaptive sampling step, which reflects the total number of PDE collocation points remains constant and does not increase with the number of iterations. The release strategy is based on the consideration that the currently selected collocation points may not necessarily need to be paid attention to in subsequent training. Thus, it helps to avoid the cumulative effects caused by previously chosen collocation points and also serves to reduce training costs.

The other is that the criterion for selecting points. We choose collocation points with the largest weighted residuals $w_i^p*\mathcal{L}_r\left(t_i, \boldsymbol{x}, \boldsymbol{\theta}\right)$, where
\begin{align}\label{L_r(t,x)}
\mathcal{L}_r\left(t, \boldsymbol{x}, \boldsymbol{\theta}\right)=\left|\frac{\partial \boldsymbol{u}_{\boldsymbol{\theta}}}{\partial t}\left(t, \boldsymbol{x}\right)+\mathcal{N}\left[\boldsymbol{u}_{\boldsymbol{\theta}}\right]\left(t, \boldsymbol{x}\right)\right|^2.
\end{align}
Here, $p \geq 0$ is a hyper-parameter. Note that if we disregard the increase in the number of collocation points with iterations in the RAR method, RAR is a special case of the current adaptive sampling strategy by choosing $p=0$. As the value of $p$ approaches 0, more points with large PDE residuals are sampled, which tend to be concentrated in the later time regions. Conversely, as the value of $p$ increases, the emphasis shifts towards weights, resulting in sampling points that are more concentrated in areas that have already been well-trained but still have relatively large PDE residuals. We consider two different approaches to determine $p$: (i) fixing it as a constant coefficient, and (ii) updating it according to a specific criterion.

The value of $p$ can control whether the weighted residuals are dominated by weights or PDE residuals, thereby controlling the  distribution of adaptive collocation points. Since the causality-guided adaptive sampling method aims to ensure that newly added sampling points are predominantly located in the region currently undergoing training, the second approach for determining $p$ should ideally update $p$ towards achieving this goal. Specifically, we propose the following update scheme for $p$ with $p_0$ as the initial value, referred to as the temporal alignment driven update (TADU) scheme:
\begin{align}\label{update_p}
	p_{n+1}=p_n\left[\beta_1 \tanh\left(\beta_2 \left(\frac{t_{ada}}{t_w}-1\right)\right)+1\right], \quad n=0,1,\cdots
\end{align}
where the parameter $\beta_1 \in (0,1)$ regulates the maximum change factor of $p$, and $\beta_2>0$ controls the difficulty level of achieving the maximum change factor. In addition, $t_{ada}$ indicates the time at which the previous batch of adaptive collocation points is concentrated, which can be determined by averaging the time coordinates corresponding to these $N_A$ newly added configuration points. Then we opt to define $t_w$ using the threshold method, which signifies the approximate time during which the model is being trained. Given a threshold $\kappa \in [0.5,1)$, $t_w$ is defined as
\begin{align}
t_w= \begin{cases}\min \{t_i: w_i< \kappa \} & \text { if } w_{N_t}< \kappa, \\ T_1 & \text { if } w_{N_t}\geq \kappa \end{cases}	
\end{align}
This scheme updates the parameter $p$ based on the relationship between $t_{\text{ada}}$ and $t_{\text{w}}$. In turn, the purpose of updating $p$ is to make these two moments as close or synchronized as possible. The initial value $p_0$ can be chosen as any constant such that $p_0 \geq 0$. The optimal value of $p$ is problem-dependent, and $p=1$ is a common and effective default choice according to the numerical results.

\begin{remark}
	When employing the temporal alignment driven update (TADU) scheme, the time interval $[T_0, T_1]$ must satisfy $T_0 \geq 0$ to ensure that $t_{ada}$ and $t_{w}$ have the same sign. This condition is easily met for evolution equations by leveraging the time translation invariance of the solution.
\end{remark}

In the initial formulation of Causal PINN, only the scenario with periodic boundary conditions (BCs) were considered, which can be strictly imposed by embedding the input coordinates into a Fourier expansion \cite{exactlyperiodicBC}. We extend this framework to the case of non-periodic boundary conditions here. For both scenarios, we re-formulate the loss functions to incorporate the causality-guided adaptive sampling method within the Causal PINN framework.

\begin{itemize}
 \item \textbf{Periodic BCs}: 

The initial set of residual points is $\mathcal{T}_f = \{(t_i,\boldsymbol{x}_j)| i=1,2,\cdots,N_t; j=1,2,\cdots,N_x\}$ and the number of collocation points at each time step is $N_x$. Every $K$ iterations, $N_A$ points with the largest weighted PDE residuals $w_i^p*\mathcal{L}_r\left(t_i, \boldsymbol{x}, \boldsymbol{\theta}\right)$ are selected from a randomly chosen dense set $U=\{(t_i, \boldsymbol{x})|i=1,2, \ldots, N_t, \boldsymbol{x} \in \boldsymbol{\Omega} \}$. Note that in the next adaptive sampling step, the $N_A$ collocation points selected this time should be released. We denote the set of newly selected collocation points as $\mathcal{T}_A$ and the set of all collocation points as $\mathcal{T}_{total}=\mathcal{T}_f \bigcup \mathcal{T}_A = \bigcup\limits_{i=1}^{N_t} \tau_i$, where $\tau_i$ represents the set of collocation points at $t = t_i$. Therefore, the points in the set $\tau_i (i=1,2,\cdots, N_t)$ are updated, potentially changing their counts $|\tau_i|$, and the loss function is modified as follows:
\begin{align}\label{periodicBC-loss}
\mathcal{L}(\boldsymbol{\theta})=\lambda_{i c} \mathcal{L}_{i c}(\boldsymbol{\theta})+\lambda_r \mathcal{L}_r(\boldsymbol{\theta}),
\end{align}
where
\begin{align}\label{L_ic}
\mathcal{L}_{i c}(\boldsymbol{\theta})=\frac{1}{N_{i c}} \sum_{i=1}^{N_{i c}}\left|\boldsymbol{u}_{\boldsymbol{\theta}}\left(T_0, \boldsymbol{x}_{i c}^i\right)-\boldsymbol{g}\left(\boldsymbol{x}_{i c}^i\right)\right|^2,	
\end{align}
\begin{align}\label{L_r}
\mathcal{L}_r(\boldsymbol{\theta})&=\frac{1}{\sum_{i=1}^{N_t} |\tau_i|} 	\sum_{i=1}^{N_t} \sum_{j=1}^{|\tau_i|} w_i \mathcal{L}_r\left(t_i, \boldsymbol{x}_{i,j}, \boldsymbol{\theta}\right)\\
&=\frac{1}{N_x\cdot N_t +N_A}  \sum_{i=1}^{N_t} \sum_{j=1}^{|\tau_i|} w_i \mathcal{L}_r\left(t_i, \boldsymbol{x}_{i,j}, \boldsymbol{\theta}\right),
\end{align}
\begin{align}\label{w_i}
w_i=\exp \left(-\epsilon \frac{1}{N_x} \sum_{k=1}^{i-1} \sum_{j=1}^{|\tau_k|} \mathcal{L}_r\left(t_k, \boldsymbol{x}_{k,j}, \boldsymbol{\theta}\right)\right), \text { for } i=2,3, \ldots, N_t,
\end{align}
and $\mathcal{L}_r\left(t, \boldsymbol{x}, \boldsymbol{\theta}\right)$ defined in Eq. \eqref{L_r(t,x)} denotes PDE residuals. Repeat this procedure until the total number of iterations reaches the limit.

\begin{remark}
The coefficient in the weight $w_i$ defined in Eq. \eqref{w_i} is taken as $\frac{1}{N_x}$ to preserve dimensional consistency while avoiding the 'dilution' of the contribution of PDE residuals at adaptive collocation points that would result from using the coefficient $\frac{1}{|\tau_k|}$.
\end{remark}

\begin{remark}
Comparison of the re-formulated and original weights: Let $S_i$ denote the number of adaptive collocation points before $t=t_i$, and then Eq. \eqref{w_i} can be expressed as follows:
\begin{align}
w_i=\exp \left(-\epsilon \sum_{k=1}^{i-1} \mathcal{L}_r\left(t_k, \boldsymbol{\theta}\right) -\epsilon \frac{1}{N_x} \sum_{n=1}^{S_i} \mathcal{L}_r\left(t_{k_n}, \boldsymbol{x}_{k_n,j_n}, \boldsymbol{\theta}\right) \right),\quad k_n \in \{1,2,\cdots,i-1 \}. 
\end{align}	
Its exponential part incorporates the PDE residuals of the new collocation points compared to Eq. \eqref{original_w_i}. This implies that the value of modified $w_i$, resulting from the introduction of the adaptive sampling strategy, is smaller than the original ones. Consequently, the network is compelled to meet higher accuracy requirements for these $S_i$ points in the earlier time regions.
\end{remark}
\end{itemize}

\begin{itemize}
 \item \textbf{Non-periodic BCs}:

In this case, it is not feasible to accurately impose boundary conditions as hard constraints, and the causal training should be adapted to incorporate boundary constraints. Utilize the following 1+1 dimensional equation as an illustrative example, 
\begin{align}
\boldsymbol{u}_t+\mathcal{N}[\boldsymbol{u}]=0, \quad t \in[T_0, T_1], \quad x \in [X_0, X_1].
\end{align}
Given the set of boundary points $\{(t_i,x_{bc})|x_{bc}= X_0\, \text{or}\, X_1\}_{i=1}^{N_t}$, we provide a re-formulation of the loss function that incorporates boundary constraints into the computation of weights:
\begin{align}
\mathcal{L}(\boldsymbol{\theta})=\lambda_{i c} \mathcal{L}_{i c}(\boldsymbol{\theta})+\lambda_{b c} \mathcal{L}_{b c}(\boldsymbol{\theta})+\lambda_r \mathcal{L}_r(\boldsymbol{\theta}),
\end{align}
where
\begin{align}
\mathcal{L}_{i c}(\boldsymbol{\theta})=\frac{1}{N_{i c}} \sum_{i=1}^{N_{i c}}\left|\boldsymbol{u}_{\boldsymbol{\theta}}\left(T_0, x_{i c}^i\right)-\boldsymbol{g}\left(x_{i c}^i\right)\right|^2,	
\end{align}
\begin{align}
\mathcal{L}_{b c}(\boldsymbol{\theta})=\frac{1}{2\cdot N_t}\sum_{i=1}^{N_t} w_i \left( \mathcal{L}_{b c}(t_i, X_0, \boldsymbol{\theta})+\mathcal{L}_{b c}(t_i, X_1, \boldsymbol{\theta}) \right),
\end{align}
\begin{align}
\mathcal{L}_{r}(\boldsymbol{\theta})	= \frac{1}{N_x\cdot N_t +N_A} \sum_{i=1}^{N_t} \sum_{j=1}^{|\tau_i|} w_i \mathcal{L}_r\left(t_i, x_{i,j}, \boldsymbol{\theta}\right),
\end{align}
\begin{align}
w_i=\exp \left(-\epsilon	 \sum_{k=1}^{i-1}  \left(\frac{1}{N_x} \sum_{j=1}^{|\tau_k|} \mathcal{L}_r\left(t_k, x_{k,j}, \boldsymbol{\theta}\right)+ \frac{1}{2} \sum_{x \in \{X_0, X_1\}} \mathcal{L}_{b c} \left( t_k, x, \boldsymbol{\theta} \right) \right)   \right),
\end{align}
with $\mathcal{L}_{b c} \left( t_i, x_{bc}, \boldsymbol{\theta} \right)=\left| \boldsymbol{u}_{\boldsymbol{\theta}}(t_i, x_{b c})-\boldsymbol{u}_{bc}^i \right|^2, x_{bc} \in \{X_0, X_1 \}$.

The aforementioned formula can also be readily extended to higher-dimensional scenarios.

\end{itemize}

\begin{remark}
	The study \cite{improvedCausal} pointed out that the causal training method does not scale for time grids with the growth of number of nodes $N_t$. To mitigate this drawback, we can use the time-marching strategy \cite{AdaptivePINN,bcPINN} and employ the causality-guided adaptive sampling method within each time window.
\end{remark}

Algorithm \ref{table2-1} summarizes the Causal AS method for the scenario with periodic boundary conditions. It employs the temporal alignment driven update scheme for $p$, the annealing scheme for $\epsilon$, and an effective stopping criterion for assessing model convergence.

\begin{algorithm}[htbp]
\SetAlgoLined  
\caption{Causal PINNs with causality-guided adaptive sampling method}
\label{table2-1} 

Consider a physics-informed neural network \( \boldsymbol{u}_{\boldsymbol{\theta}}(t, \boldsymbol{x}) \) imposed the exact boundary conditions, and the corresponding weighted loss function
\begin{align}  
\mathcal{L}(\boldsymbol{\theta})=\lambda_{i c} \mathcal{L}_{i c}(\boldsymbol{\theta})+\lambda_r \mathcal{L}_r(\boldsymbol{\theta}), \nonumber
\end{align} 
where \( \mathcal{L}_{i c}(\boldsymbol{\theta}) \) and \( \mathcal{L}_r(\boldsymbol{\theta}) \) are defined in Eq.\eqref{L_ic} and Eq.\eqref{L_r}. Initialize \( w_1 \) by 1 and select an increasing sequence of the causality parameter \( \{\epsilon_i\}_{i=1}^k \).  The initial set of residual points is $\mathcal{T}_f$. Specify the initial value $p_0$, the threshold $\kappa$ and the parameters $\beta_1$ and $\beta_2$.

Then use \( S \) steps of a gradient descent algorithm to update the parameters \( \boldsymbol{\theta} \) as:

\For{\( \epsilon = \epsilon_1, \ldots, \epsilon_k \)}{
    \For{\( l = 1, \ldots, S \)}{
        (a) Compute and update the temporal weights by
        \begin{align} 
        w_i=\exp \left(-\epsilon \frac{1}{N_x} \sum_{k=1}^{i-1} \sum_{j=1}^{|\tau_k|} \mathcal{L}_r\left(t_k, \boldsymbol{x}_{k,j}, \boldsymbol{\theta}\right)\right), \text { for } i=2,3, \ldots, N_t.\nonumber
        \end{align}  
        Here \( \epsilon > 0 \) is a user-defined hyper-parameter that determines the "slope" of temporal weights. \\
        (b) Update the parameters \( \boldsymbol{\theta} \) via gradient descent
        \begin{align}  
        \boldsymbol{\theta}_{l+1} = \boldsymbol{\theta}_l - \nabla \mathcal{L}(\boldsymbol{\theta}_l).\nonumber
        \end{align}  
        \If{\( \min_i w_i > \delta \)}{
            break;
        }
        \If{\( l \bmod K == 0 \)}{
        	  (a) Compute \(t_{ada} \) and \(t_{w} \) to  update $p$
        	  \begin{align} 
        	  	p_{n+1}=p_n\left[\beta_1 \tanh\left(\beta_2 \left(\frac{t_{ada}}{t_w}-1\right)\right)+1\right].\nonumber
        	  \end{align}
        	  
        	  (b) Release the set of adaptive collocation points \( \mathcal{T}_A \).\\
              (c) Randomly sample a set of dense points \( U=\{(t_i, \boldsymbol{x})|i=1,2, \ldots, N_t, \boldsymbol{x} \in \boldsymbol{\Omega} \} \).\\
              (d) Compute the weighted PDE residuals $w_i^{p_{n+1}}*\mathcal{L}_r\left(t_i, \boldsymbol{x}, \boldsymbol{\theta}\right)$ for the points in \( U \) and select \( N_A \) points with the largest weighted residuals to update the set \( \mathcal{T}_A \). \\
              (e) Update the set of all collocation points \( \mathcal{T}_{total}=\bigcup\limits_{i=1}^{N_t} \tau_i\) and the set of residual points \( \tau_i \) at each time point $t_i$.
           
        }
    }
}

The recommended hyper-parameters are \( \delta = 0.99 \) and \( \{\epsilon_i\}_{i=1}^k = [10^{-2}, 10^{-1}, 10^0, 10^1, 10^2] \).
\end{algorithm}

\subsubsection{Related work}
\quad

A similar idea has been proposed in \cite{CausalR3}. Daw et al. first proposed a Retain-Resample-Release sampling (R3) strategy. Unlike the common approach of adding points with large PDE residuals, this study suggests removing points with small PDE residuals. Besides, it incorporates new points through uniform random sampling instead of introducing new points with large PDE residuals. Subsequently, a causal formulation of PDE loss was introduced, which achieves a similar effect to Causal PINN \cite{CausalPINN} in respecting the principle of causality. In Causal PINN, causality is enforced using the time-dependent weight $w(t)$ directly characterized by the PDE residuals while in Ref. \cite{CausalR3}, the gate function $g(t)$ is constructed to serve a similar purpose. The updatable parameter $\gamma$ in the gate function controls the portion of time revealed to the model and will gradually increase to reveal more portions of the time domain when the PDE residual is sufficiently low. Building on the R3 sampling, Causal R3 sampling was subsequently proposed by modifying the residual function to account for the causal gate values.

For the causality-enforced training method proposed by Daw et al., the parameter $\gamma$ is updated based on the PDE residual values, thereby determining the gate function $g(t)$. In contrast, in the Causal PINN method proposed by Wang et al., the PDE residual values directly determine the temporal weight $w(t)$. The differences in implementing causality leads to distinctions between our proposed Causal AS method within the Causal PINN framework and the Causal R3 sampling method proposed by Daw et al. We will compare the effectiveness of these two sampling strategies in the subsequent numerical experiments section.

\section{Numerical experiments}\label{experiments}

In this section, we aim to demonstrate the effectiveness of the Causal AS method through numerical results. We consider various types of equations here, including the Allen-Cahn equation, the nonlinear Schr\"{o}dinger equation and the Korteweg-de Vries equation with periodic boundary conditions, as well as the modified Korteweg-de Vries equation with non-periodic boundary conditions. Morever, we evaluate and compare the performance of the Causal AS and Causal R3 methods under identical parameter settings.

Throughout all benchmarks, we employ the fully-connected feedforward neural networks with hyperbolic tangent ($\tanh$) activation function and the weights are initialized with Xavier initialization method. When implementing Causal R3, the initial value of the parameter $\gamma$ in the gate function is set to $-0.5$, with the parameter $\alpha$ determining the steepness of the gate  assigned a value of 5. All parameters involved in the update scheme for $\gamma$ use the default settings. All networks are trained utilizing the Adam optimizer along with an exponential learning rate decay, featuring a decay rate of 0.9 every 5000 iterations on NVIDIA Tesla P100 GPU (3584 CUDA cores and 16 GB of HBM2 vRAM). We remark that all the numerical results presented below are the best outcomes from 10 random experiments we conducted.

\subsection{Allen-Cahn equation}\label{section_AC}
\quad

The Allen-Cahn equation \cite{AC}, introduced by S.M. Allen and J.W. Cahn in 1979, is a nonlinear partial differential equation that describes phase transition processes. It is widely used to model the evolution of phase interfaces in materials science. Here, we consider the Allen-Cahn equation in the following form
\begin{align}\label{AC}
& u_t-0.0001 u_{x x}+5 u^3-5 u=0, \quad t \in[0,1], x \in[-1,1], \\
& u(0, x)=x^2 \cos (\pi x), \\
& u(t,-1)=u(t, 1), \\
& u_x(t,-1)=u_x(t, 1).
\end{align}
When using the original PINN method to solve this problem, it fails to capture even the basic structure of the solution. Subsequent variants, such as adaptive time sampling\cite{AdaptivePINN}, bc-PINN\cite{bcPINN} and SA-PINN\cite{SA-PINNs}, can improve prediction accuracy but demonstrate insufficient capability in predicting sharp local extrema at later times. However, the causal training method proposed by Wang et al.\cite{CausalPINN} effectively addresses this limitation and provides much more reliable solutions, achieving a relative $\mathbb{L}_2$ error of 1.43e-03 using the multi-layer perceptron (MLP) architecture.

Following this study, we design the causality-guided adaptive sampling method within the Causal PINN framework. For this example, it satisfies periodic boundary conditions in the spatial direction with a period $L = 2$, and then we construct Fourier features embedding of the input to impose periodic boundary conditions as hard constraints \cite{exactlyperiodicBC}
\begin{align}
e(t, x)=(t, 1, \cos(\omega x), \sin(\omega x), \cos(2\omega x), \sin(2\omega x), \cdots, \cos(m \omega x), \sin(m \omega x)),	
\end{align}
where $\omega=\frac{2 \pi}{L}$ and $m$ is a non-negative integer. Here, we take $m=10$ and it can be verified that the output of the neural network $u_{\boldsymbol{\theta}}(e(t,x))$ satisfies the periodic constraint. The training dataset and the neural network architecture are consistent with the setup in Ref. \cite{CausalPINN}. Specifically, we take $N_{ic}=256$ and generate uniform grids with sizes $N_t=100$ in the temporal domain and $N_x=256$ in the spatial domain, respectively. Every $K$ iterations, a dense set of points $U=\{(t_i, x)|i=1,2, \ldots, N_t, x \in [-1,1] \}$ is randomly generated. The weighted PDE residual is computed for each point in $U$, and the $N_A$ points with the highest values are selected to form the adaptive sampling set $\mathcal{T}_A$. The ratio $\rho=\frac{N_A}{|U|}$ represents the relationship between the quantities of points in the two sets and should be configured to an appropriate value. Note that before conducting the next adaptive sampling step, the adaptive collocation points obtained in the current iteration should be discarded, ensuring that the total number of configuration points is consistently $N_f=N_x\cdot N_t +N_A$. The loss functions are given in Eqs. \eqref{periodicBC-loss}-\eqref{w_i}. Here, we set the parameters as $K = 500, \rho = \frac{2}{5}, \lambda_{i c}=100, \lambda_r=1$ and $\epsilon = 100$. A fully-connected neural network with 4 hidden layers and 128 neurons per hidden layer is constructed to approximate the solution, utilizing the $\tanh$ activation function. Then we train the neural network for $3 \times 10^5$ iterations by using the Adam optimizer.

\begin{figure}[htbp]
\centering
\includegraphics[width=5.5cm,height=4.2cm]{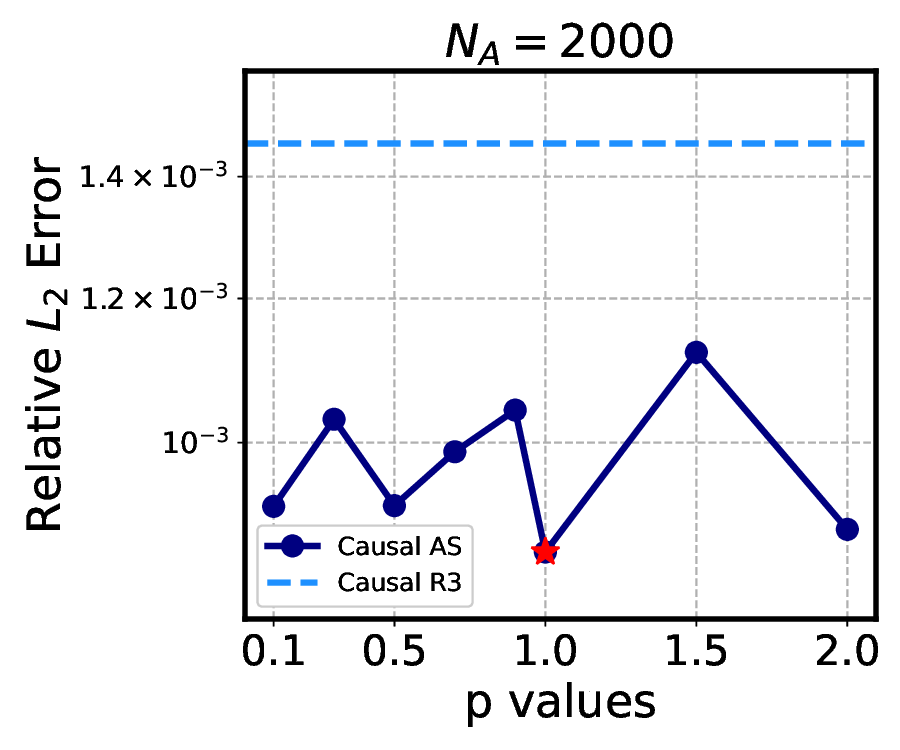}
$a$
\includegraphics[width=5.5cm,height=4.2cm]{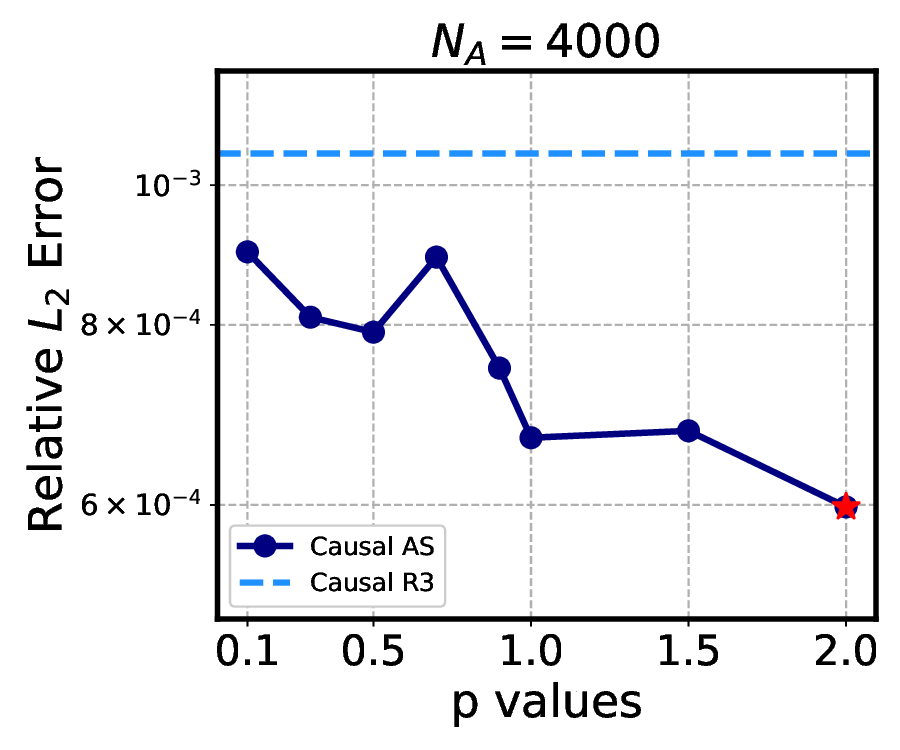}
$b$
\includegraphics[width=5.5cm,height=4.2cm]{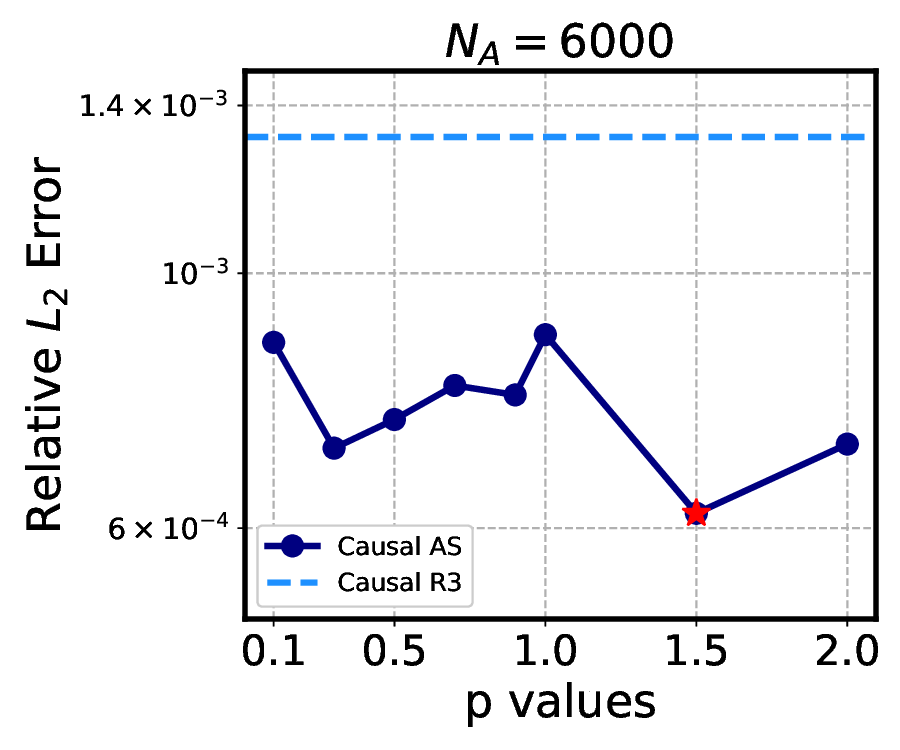}
$c$\\
\includegraphics[width=5.5cm,height=4.2cm]{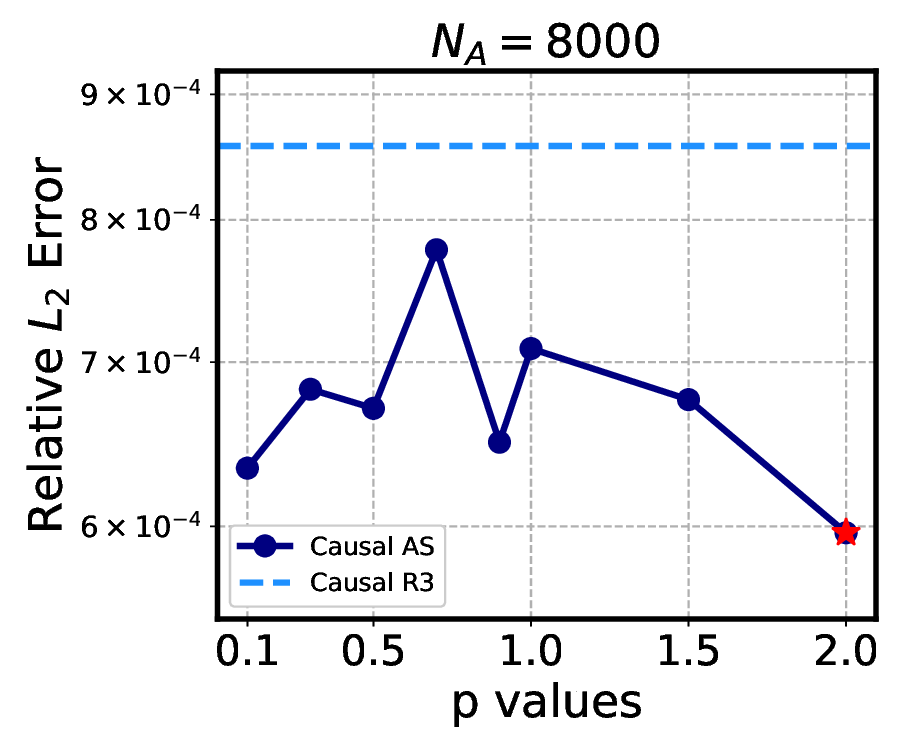}
$d$
\includegraphics[width=5.5cm,height=4.2cm]{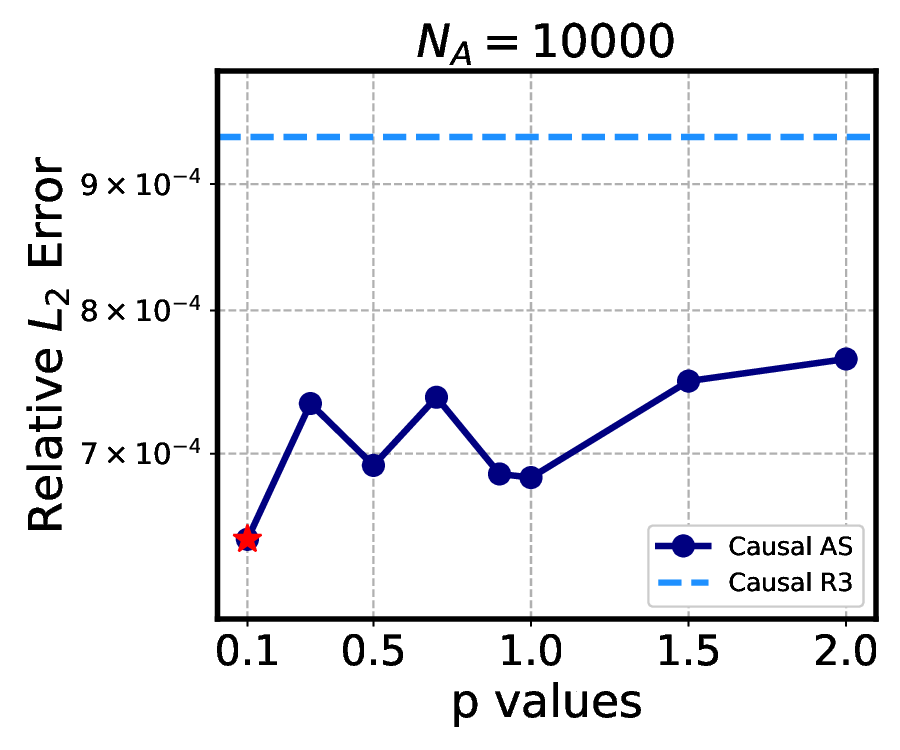}
$e$
\caption{(Color online) Comparisons of the relative $\mathbb{L}_2$ errors between the Causal R3 and Causal AS methods with different values of $p$ and $N_A$: (a) $N_A=2000$; (b) $N_A=4000$; (c) $N_A=6000$; (d) $N_A=8000$; (e) $N_A=10000$;}
\label{fig3-1}
\end{figure}

For the approach where $p$ is kept constant, we investigate the values $p = 0.1, 0.3, 0.5, 0.7, 0.9, 1, 1.5$ and  $2$. For each $p$, the number of adaptive collocation points $N_A$ is varied as $2 \times 10^3, 4 \times 10^3, 6 \times 10^3, 8 \times 10^3,$ and $1 \times 10^4$. Subsequently, the accuracy of the Causal AS method is assessed for each configuration. Additionally, while maintaining the same total number of collocation points $N_f$, the network architecture and other parameters, we also employ the Causal R3 method to numerically solve the Allen-Cahn equation. The relative $\mathbb{L}_2$ errors of the two methods, as influenced by variations in $p$ and $N_A$, are illustrated in Fig. \ref{fig3-1}. The light blue dashed line indicates the error produced by the Causal R3 method. It is evident that, for all constant values of $p$, our proposed Causal AS method consistently exhibits higher accuracy compared to the Causal R3 method.

\begin{figure}[htbp]
\centering
\includegraphics[width=16cm,height=4.2cm]{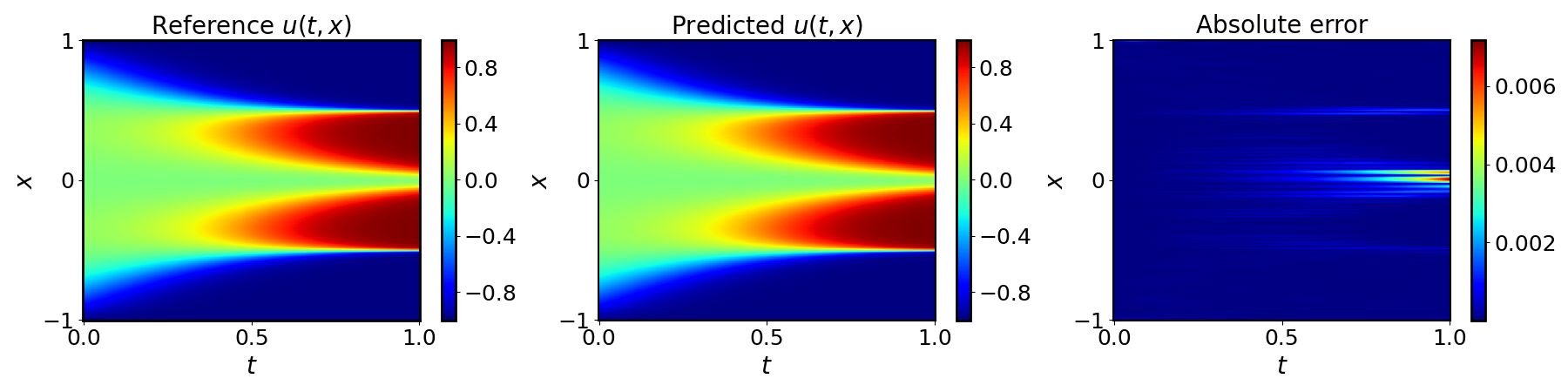}
$a$
\includegraphics[width=14cm,height=4cm]{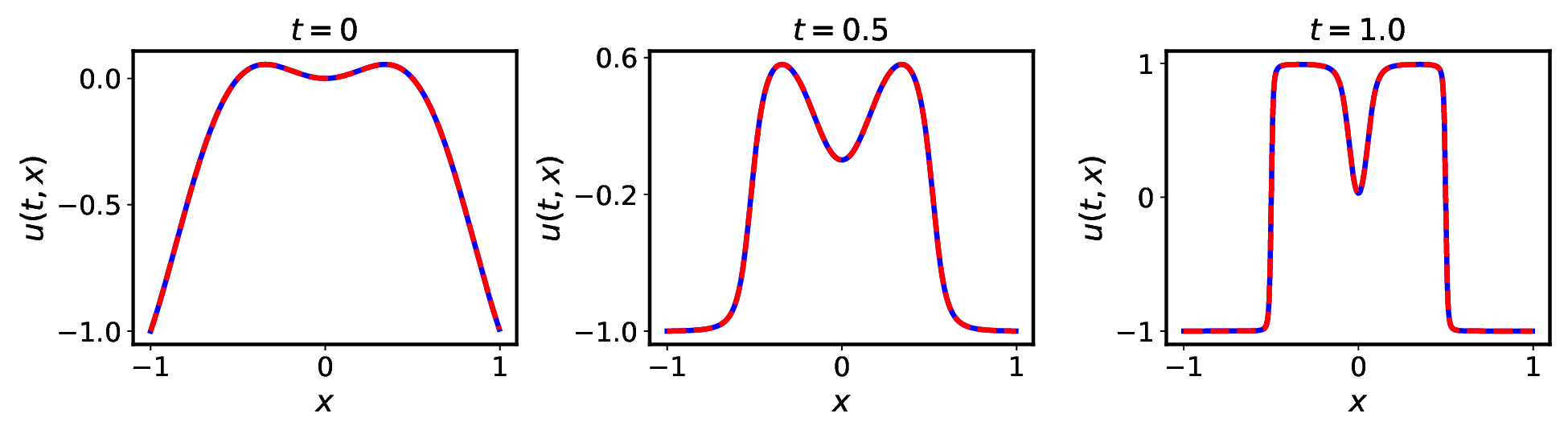}
$b$
\caption{(Color online) Allen–Cahn equation: (a) The reference solution, predicted solution obtained using the Causal AS method and the absolute error; (b) Comparison of the predicted and reference solutions corresponding to the three temporal snapshots at $t=0,0.5,1.0$.}
\label{fig3-2}
\end{figure}

\begin{figure}[htbp]
\centering
\includegraphics[width=6cm,height=4cm]{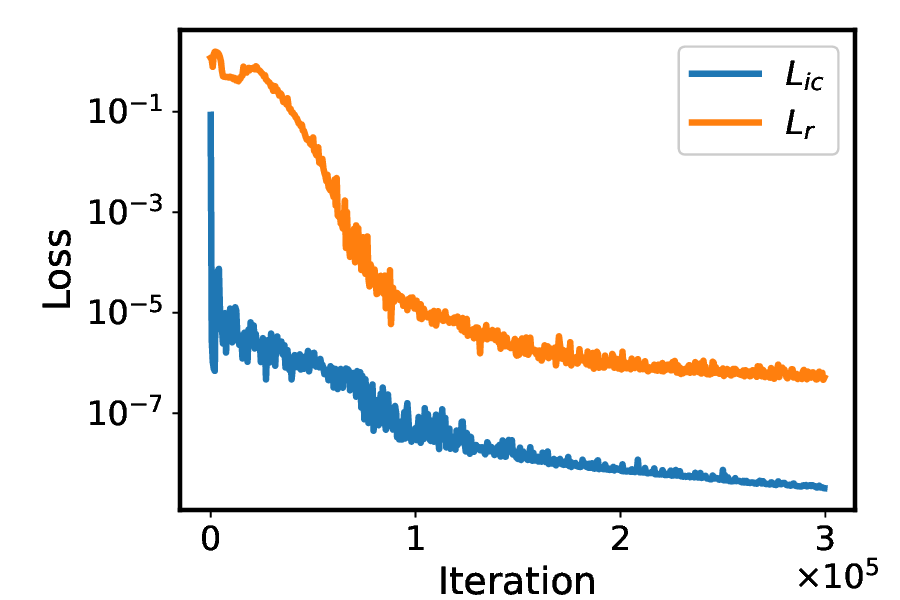}
$a$
\includegraphics[width=7cm,height=4cm]{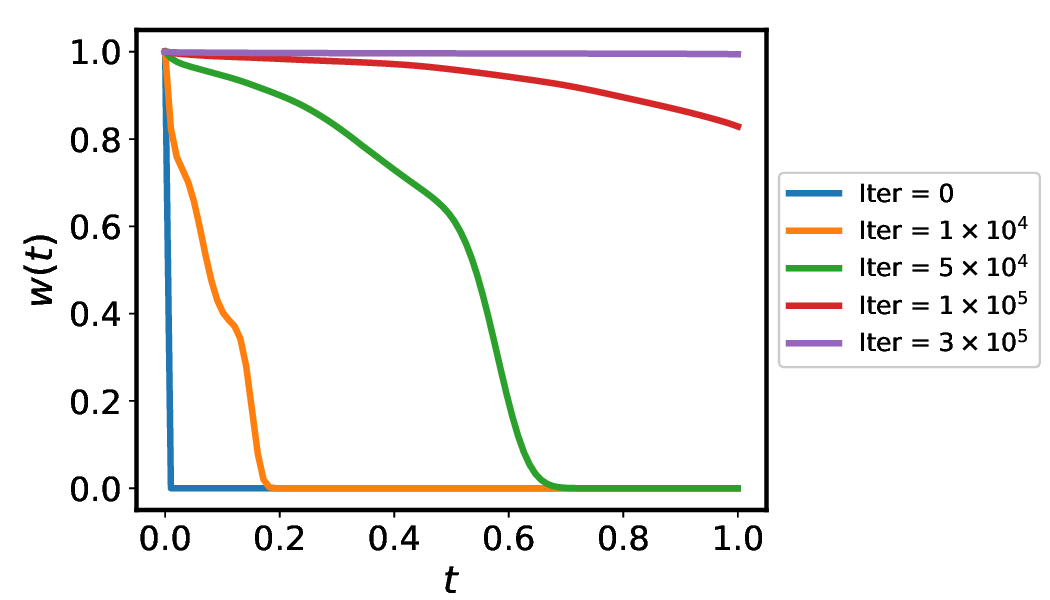}
$b$
\caption{(Color online) Allen–Cahn equation: (a) Loss convergence of Causal AS method; (b) Temporal weight $w(t)$ at different iterations of the training.}
\label{fig3-3}
\end{figure}

\begin{figure}[htbp]
\centering
\includegraphics[width=16cm,height=4.2cm]{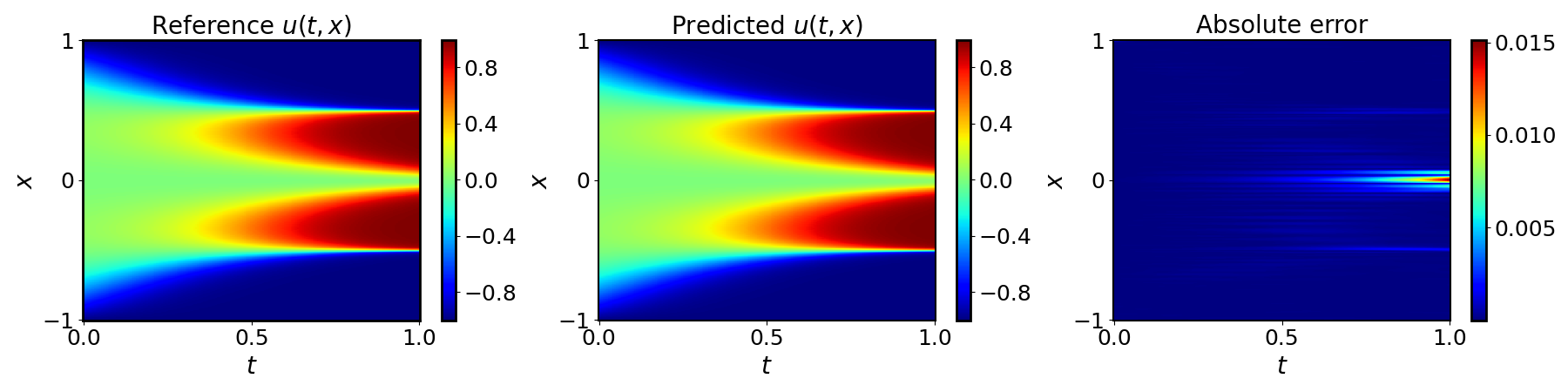}
\caption{(Color online) Allen–Cahn equation: The reference solution, predicted solution obtained using the Causal R3 method and the absolute error.}
\label{fig3-4}
\end{figure}

As an example, with $N_A = 4000$ and $p = 2$, the numerical results of the Causal AS method are shown in Fig. \ref{fig3-2} and Fig. \ref{fig3-3}, yielding a relative $\mathbb{L}_2$ error of $5.981e-04$. Fig. \ref{fig3-3} illustrates the convergence behavior of different loss terms and reveals that the weights at each time point are nearly equal to 1 by the end of the iterations. These figures clearly show a high degree of concordance between the predicted and reference solutions. In contrast, Fig. \ref{fig3-4} displays the results for the Causal R3 method under the same parameters with a resulting relative $\mathbb{L}_2$ error of $1.052e-03$. Furthermore, a comparison of the absolute errors in both figures underscores the superior accuracy of the Causal AS method. The highest accuracy and corresponding $p$ value are highlighted with red pentagrams in Fig. \ref{fig3-1}, and these results are summarized in Table \ref{table3-1}.

\begin{table}[htbp]
\caption{Relative $\mathbb{L}_2$ errors of the solution for the Allen-Cahn equation by Causal R3 and Causal AS.}
\label{table3-1} 
\centering
\begin{tabular}{cccccc}
\bottomrule
Method    & $N_A=2000$ & $N_A=4000$ & $N_A=6000$ & $N_A=8000$ & $N_A=10000$ \\ \hline
Causal R3 & 1.460e-03    & 1.052e-03     & 1.314e-03     & 8.574e-04     &  9.403e-04    \\
Causal AS & 8.708e-04($p=1$)     &5.981e-04($p=2$)     & 6.181e-04($p=1.5$)     & 5.964e-04($p=2$)     & 6.463e-04($p=0.1$)     \\ \toprule
\end{tabular}
\end{table}

\begin{figure}[htbp]
\centering
\includegraphics[width=6cm,height=4cm]{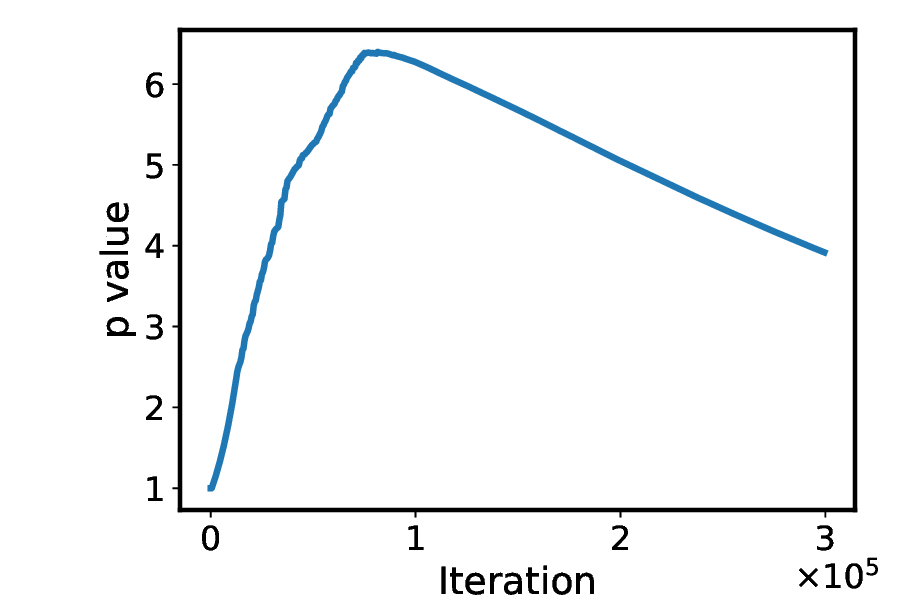}
$a$
\includegraphics[width=6cm,height=4cm]{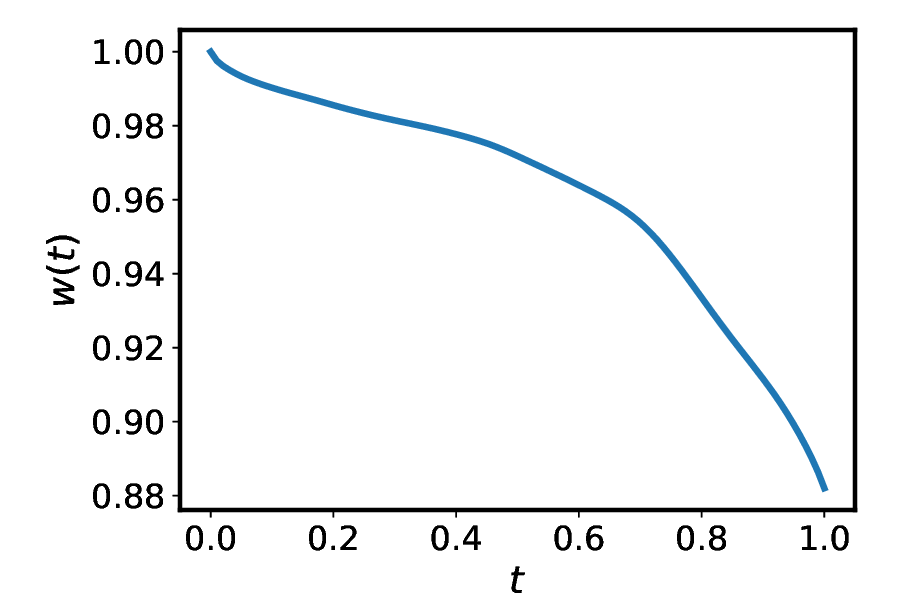}
$b$
\caption{(Color online) Allen–Cahn equation: (a) The update process of $p$; (b) Temporal weight $w(t)$ at the $1 \times 10^5$th iteration}
\label{fig3-15}
\end{figure}

We also implemented the temporal alignment driven update scheme. Using $p_0 = 1$ as the initial value, $p$ is updated every $K = 500$ iterations according to Eq. \eqref{update_p}, with parameters set to $\beta_1 = \beta_2 = 0.05$ and $\kappa = 0.9$. The number of $N_A$ also changes from 2000 to 10000 with the step size 2000. The numerical experiment results, along with those for a constant $p=1$ under the same parameters, are summarized in Table \ref{table3-2}, with smaller errors highlighted in bold. It is evident that both approaches for determining $p$ can accurately approximate the solution of the Allen-Cahn equation. Furthermore, the accuracy of the Causal AS method consistently surpasses that of the Causal R3 method regardless of the strategy used to define $p$. In addition, we illustrate the update process of $p$ for $N_A = 4000$ in Fig. \ref{fig3-15} (a). The value of $p$ exhibits an increasing trend followed by a decrease. At the start of the training, the region indicated by $t_w$ is positioned towards the front. However, due to the selection setting, the adaptive collocation points cannot concentrate on a very small $t_{ada}$ value. As a result, the TADU strategy shifts the sampling distribution forward, causing the initial increase in $p$. Around the $1 \times 10^5$th iteration, $p$ starts to decrease. The temporal weight at this stage is shown in Fig. \ref{fig3-15} (b), which indicates that the smallest $w_i $ is close to the selected $\kappa$. It is anticipated that soon after, $t_w$ will reach 1. To ensure the adaptive collocation points progressively shift backward, $p$ is continuously reduced.

\begin{table}[htbp]
\caption{Relative $\mathbb{L}_2$ errors of the solution for the Allen-Cahn equation by Causal AS using different approaches to determine $p$.}
\label{table3-2} 
\centering
\begin{tabular}{cccccc}
\bottomrule
Method    & $N_A=2000$ & $N_A=4000$ & $N_A=6000$ & $N_A=8000$ & $N_A=10000$ \\ \hline
$p\equiv 1$ & \pmb{8.708e-04}    & 6.680e-04     & 8.842e-04    & 7.090e-04    & \pmb{6.846e-04}     \\
Updating $p$ ($p_0=1$) & 1.032e-03     &\pmb{6.494e-04}     &\pmb{7.124e-04}      & \pmb{6.754e-04}     & 7.282e-04     \\ \toprule
\end{tabular}
\end{table}

\begin{figure}[htbp]
\centering
\includegraphics[width=5.5cm,height=4cm]{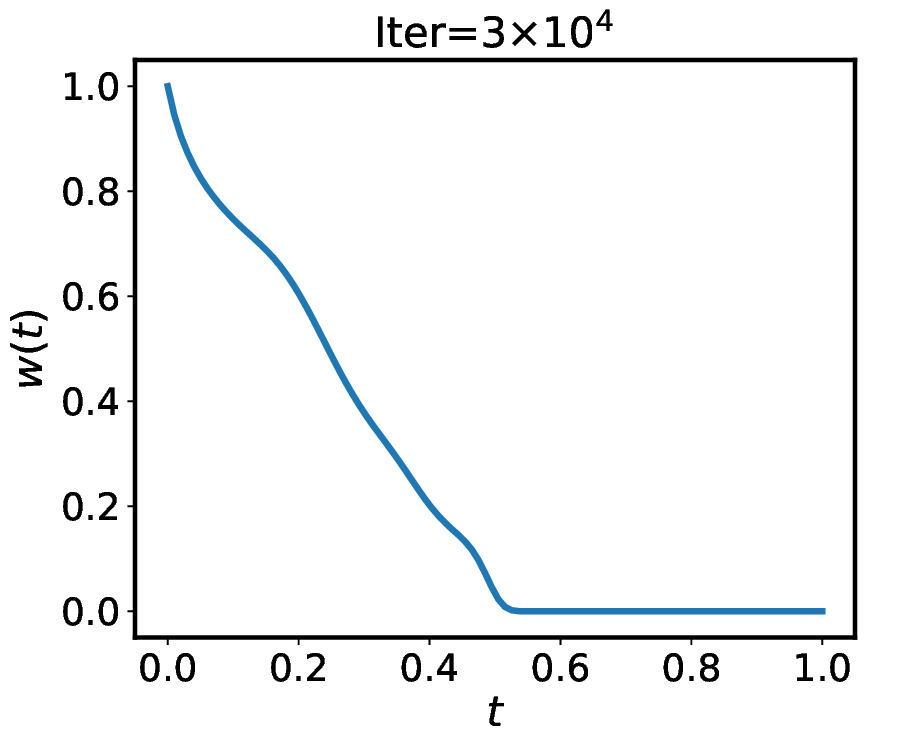}
$a$
\includegraphics[width=5.5cm,height=4cm]{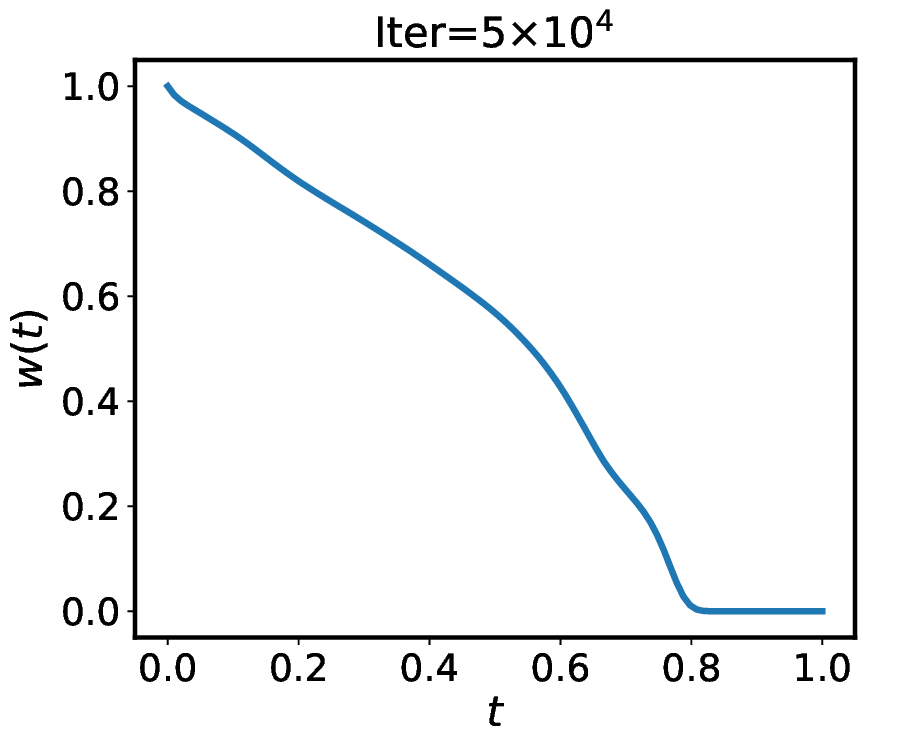}
$b$
\includegraphics[width=5.5cm,height=4cm]{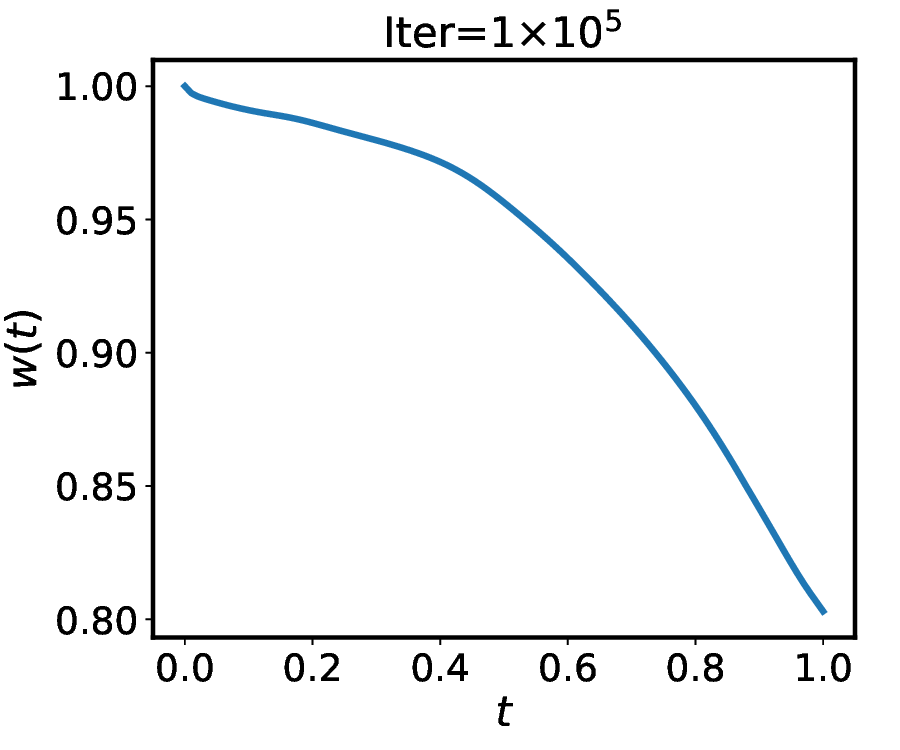}
$c$\\
\includegraphics[width=5.5cm,height=3.8cm]{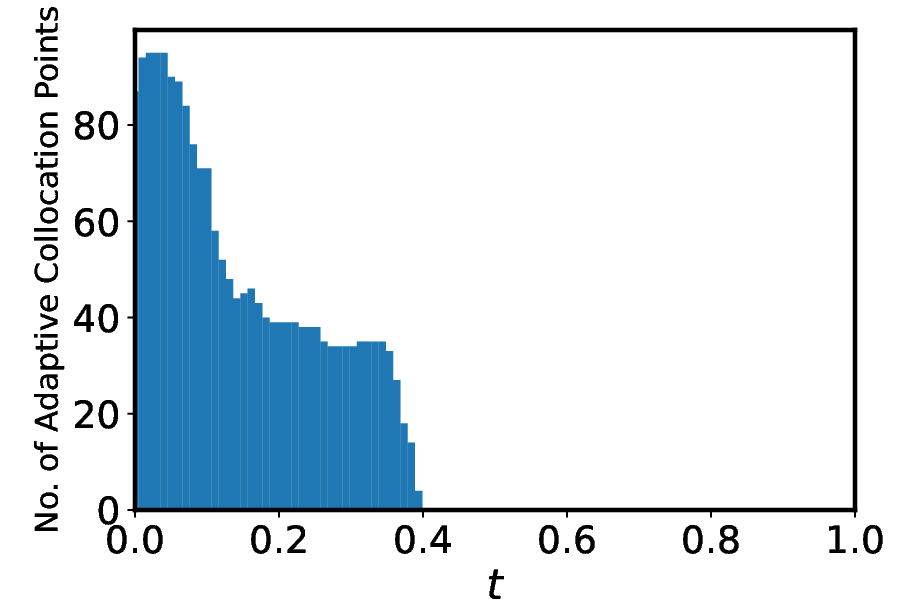}
$d$
\includegraphics[width=5.5cm,height=3.8cm]{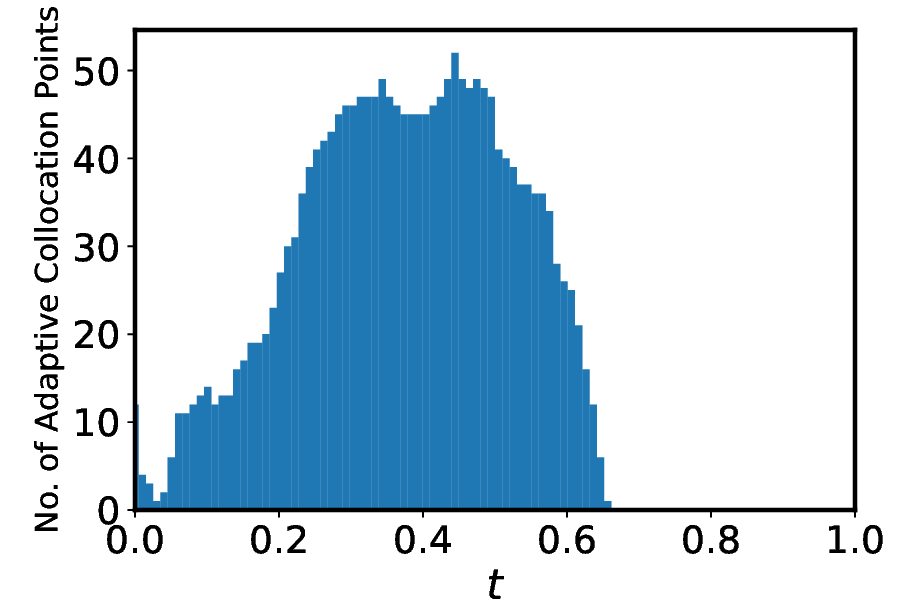}
$e$
\includegraphics[width=5.5cm,height=3.8cm]{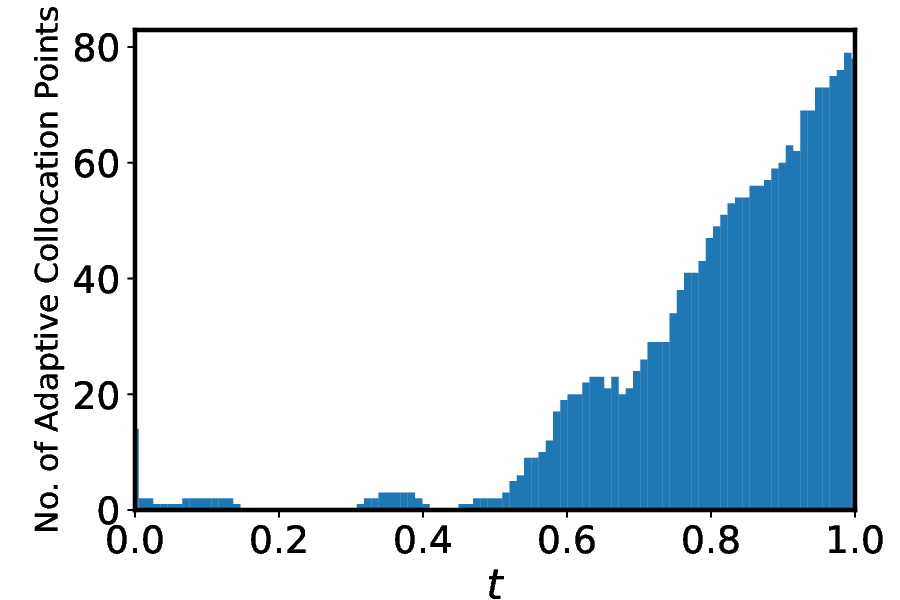}
$f$
\caption{(Color online) Temporal weight $w(t)$ and the distribution of adaptive collocation points by Causal AS method : Left: At the $3 \times 10^4$th iteration; Middle: At the $5 \times 10^4$th iteration; Right: At the $1 \times 10^5$th iteration.}
\label{fig3-5}
\end{figure}

To explore whether the point selection of the Causal AS algorithm align with our desired goals, we used the parameter settings of $N_A = 4000$ and $p = 1$ as an example. Fig. \ref{fig3-5} depicts the weight function $w(t)$ and the distribution of the 2000 sampling points with the highest weighted PDE residuals at iterations $3 \times 10^4$, $5 \times 10^4,$ and $1 \times 10^5$. This figure demonstrates that the second type of region mentioned in Section \ref{CausalAS} receives significant attention, indicating that the proposed adaptive sampling strategy is both effective and perfectly aligned with its intended motivation.

\begin{figure}[htbp]
\centering
\includegraphics[width=10cm,height=6cm]{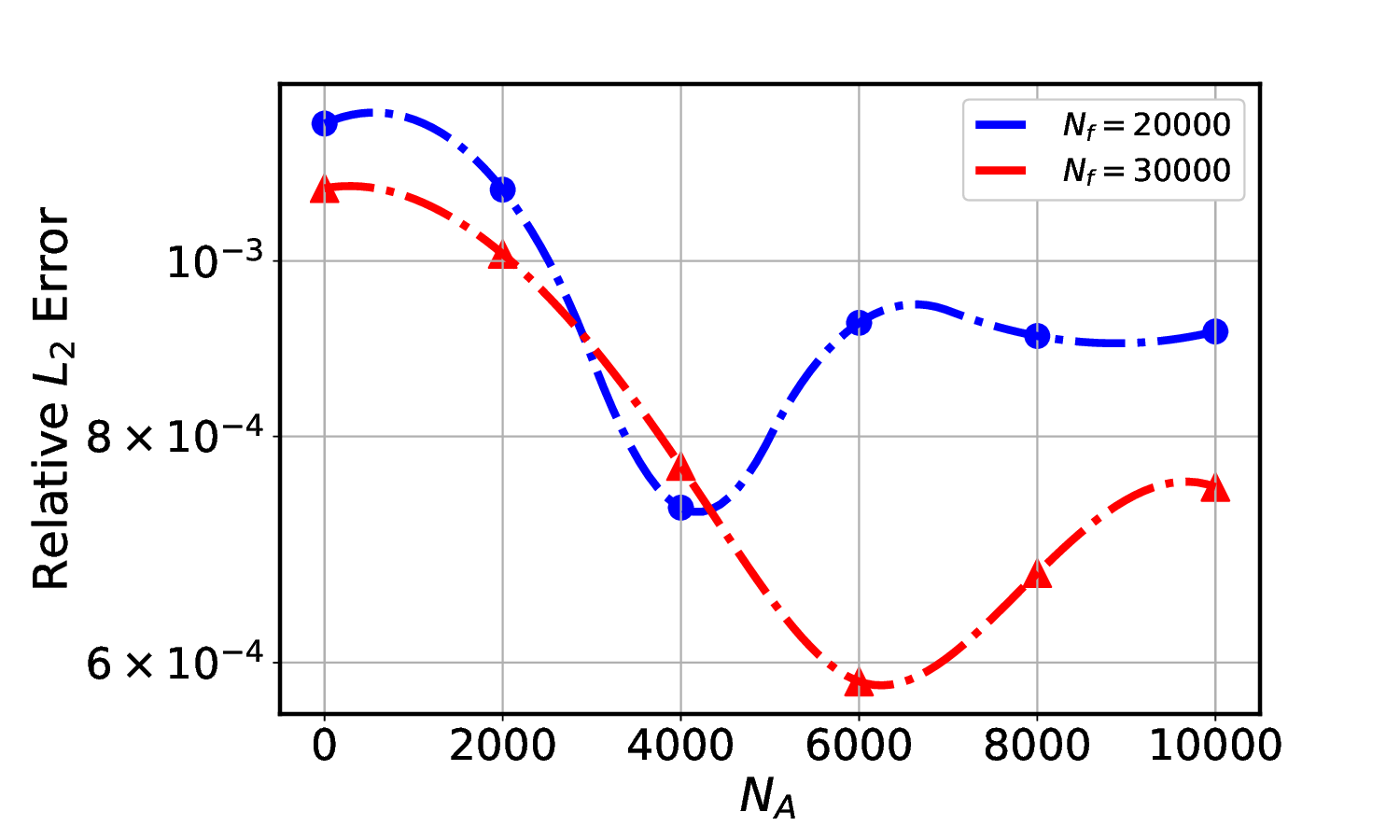}
\caption{(Color online) Allen–Cahn equation: Variation of relative $\mathbb{L}_2$ errors with respect to the number of adaptive collocation points $N_A$ under the constant total number of configuration points $N_f$}
\label{fig3-6}
\end{figure}

Given that the configuration points in the Causal AS method are composed of $\mathcal{T}_f$ and $\mathcal{T}_A$, with quantities $N_x \times N_t$ and $N_A$ respectively, we investigate how varying the number of initial collocation points and adaptive collocation points affects the outcomes, while maintaining a fixed total number of configuration points $N_f$. Here, we examined two scenarios: $N_f = 20000$ and $N_f = 30000$ . For each case, the number of adaptive collocation points $N_A$ was varied from 0 to 10000 in increments of 2000. It is worth noting that when $N_A=0$, it corresponds to the original Causal PINN method. Additionally, $N_t$ is fixed at 100, while $N_x$ varies with the number of adaptive collocation points $N_A$. The results of our numerical experiments are illustrated in Fig. \ref{fig3-6}. We used spline interpolation to generate smooth curves for the two sets of results, which can better characterize the data trends. It is evident that, with a constant total number of configuration points $N_f$, the relative $\mathbb{L}_2$ error initially decreases and then increases as $N_A$ increases. This suggests that simply increasing the number of adaptive collocation points does not necessarily lead to better results; the initial configuration points are equally important as they ensure robust training of the Causal PINN. Hence, maintaining an appropriate balance in the quantities of initial collocation points and adaptive collocation points is crucial for maximizing the effectiveness of the Causal AS method.

\subsection{Nonlinear Schr\"{o}dinger equation}\label{section_NLS}
\quad

The nonlinear Schr\"{o}dinger (NLS) equation stands as a vital model in both quantum mechanics and nonlinear optics, adept at capturing the intricate wave propagation phenomena within complex systems. This equation extends the classical Schr\"{o}dinger equation by incorporating nonlinear effects, thereby offering a more comprehensive framework for understanding such dynamic behaviors. The NLS equation is extensively utilized across multiple disciplines, serving to elucidate the propagation of light pulses through nonlinear media in fiber optics \cite{Zakharov1972}, characterize the evolution of atomic wave functions within Bose-Einstein condensates in physics \cite{Pitaevskii1961}, and capture the complex nonlinear phenomena observed in deep water waves in fluid dynamics \cite{Zakharov1968}. It is also a fundamental model in integrable systems, featuring various soliton \cite{Matveev1991}, breather \cite{NLSbreather} and rogue wave solutions \cite{NLSroguewave}.

Next, we study the following form of the NLS equation in the computational domain $t \in[0,\frac{\pi}{2}], x \in[-5,5]$:
\begin{align}
\psi_t=0.5 i \psi_{x x}+i|\psi|^2 \psi, 	
\end{align}
subject to the initial and periodic boundary conditions
\begin{align}
&\psi(0,x)=2 \operatorname{sech}(x), \\
& \psi(t,-5)=\psi(t, 5), \\
& \psi_x(t,-5)=\psi_x(t, 5).
\end{align}
To evaluate the performance of our method, we use the Fourier pseudo-spectral method to obtain the reference solution by using the Chebfun package \cite{Chebfun}. Specifically, we employ a spectral Fourier discretization with 512 modes in the spatial domain, along with a fourth-order explicit Runge-Kutta temporal integrator using a time step of $\Delta t = \frac{\pi}{2} \cdot 10^{-6}$. Based on the resulting training data, we constructed a neural network with 3 hidden layers, each containing 128 neurons, to numerically solve the NLS equation. This network has two outputs, which respectively approximate  the real part $u(t,x)$ and the imaginary part $v(t,x)$ of the solution $\psi(t,x)$. In this example, we similarly strictly enforce the periodic boundary conditions as hard constraints, setting the parameter $m$ to 10. Then the PDE residual is modified to:
\begin{align}
\mathcal{L}_r\left(t, x, \boldsymbol{\theta}\right)=\left|\frac{\partial u_{\boldsymbol{\theta}}}{\partial t}\left(t, x\right)+ 0.5 \frac{\partial^2 v_{\boldsymbol{\theta}}}{\partial x^2}\left(t, x\right)+ \left(u_{\boldsymbol{\theta}}^2+v_{\boldsymbol{\theta}}^2 \right)v_{\boldsymbol{\theta}} \right|^2 + \left|\frac{\partial v_{\boldsymbol{\theta}}}{\partial t}\left(t, x\right)- 0.5 \frac{\partial^2 u_{\boldsymbol{\theta}}}{\partial x^2}\left(t, x\right)- \left(u_{\boldsymbol{\theta}}^2+v_{\boldsymbol{\theta}}^2 \right)u_{\boldsymbol{\theta}} \right|^2.
\end{align}
and we can define the following loss function
\begin{align}
\mathcal{L}(\boldsymbol{\theta})=\lambda_{i c} \mathcal{L}_{i c}(\boldsymbol{\theta})+\lambda_r \mathcal{L}_r(\boldsymbol{\theta}),
\end{align}
where
\begin{align}
\mathcal{L}_{i c}(\boldsymbol{\theta})=\frac{1}{N_{i c}} \sum_{i=1}^{N_{i c}}\left( \left|u_{\boldsymbol{\theta}}\left(0, x_{i c}^i\right)-2 \operatorname{sech}\left(x_{i c}^i\right)\right|^2 + \left|v_{\boldsymbol{\theta}}\left(0, x_{i c}^i\right)\right|^2 \right),	
\end{align}
\begin{align}
\mathcal{L}_r(\boldsymbol{\theta})&=\frac{1}{\sum_{i=1}^{N_t} |\tau_i|} 	\sum_{i=1}^{N_t} \sum_{j=1}^{|\tau_i|} w_i \mathcal{L}_r\left(t_i, x_{i,j}, \boldsymbol{\theta}\right)\\
&=\frac{1}{N_x\cdot N_t +N_A}  \sum_{i=1}^{N_t} \sum_{j=1}^{|\tau_i|} w_i \mathcal{L}_r\left(t_i, x_{i,j}, \boldsymbol{\theta}\right),
\end{align}
\begin{align}
w_i=\exp \left(-\epsilon \frac{1}{N_x} \sum_{k=1}^{i-1} \sum_{j=1}^{|\tau_k|} \mathcal{L}_r\left(t_k, x_{k,j}, \boldsymbol{\theta}\right)\right), \text { for } i=2,3, \ldots, N_t.
\end{align}
To mitigate optimization challenges, we implemented the time-marching strategy \cite{AdaptivePINN,bcPINN}. To be specific, we divide the time domain equally into 4 sub-domains and solve them sequentially. Starting from the second sub-domain, the initial conditions are determined based on the predictions of the previous neural network at the final time. Here, the parameters are set as $N_t=20, N_x=250, N_{ic}=256, \rho = \frac{1}{4}, \lambda_{i c}=100, \lambda_r=1$ and $K = 500$. Besides, the annealing strategy is employed with $\delta = 0.99$ and $\{\epsilon_i\}_{i=1}^k = [10^{-2}, 10^{-1}, 10^0, 10^1, 10^2]$. The maximum number of iterations is taken as $3 \times 10^5$ for every tolerance $\epsilon$ in each time window but the total number of iterations for each sub-domain may vary due to the stopping criterion.

\begin{figure}[htbp]
\centering
\includegraphics[width=16cm,height=4.2cm]{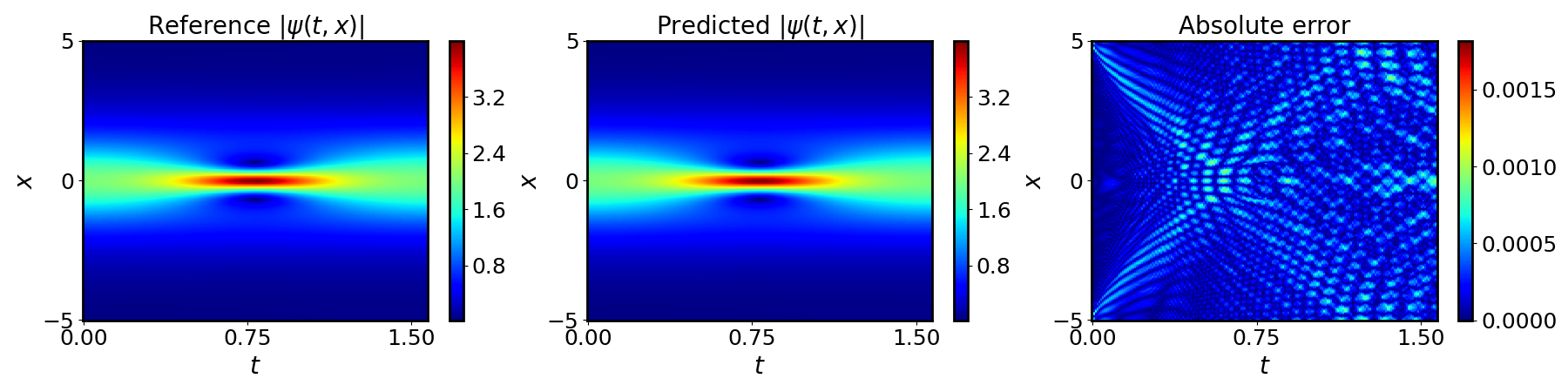}
$a$
\includegraphics[width=14cm,height=4cm]{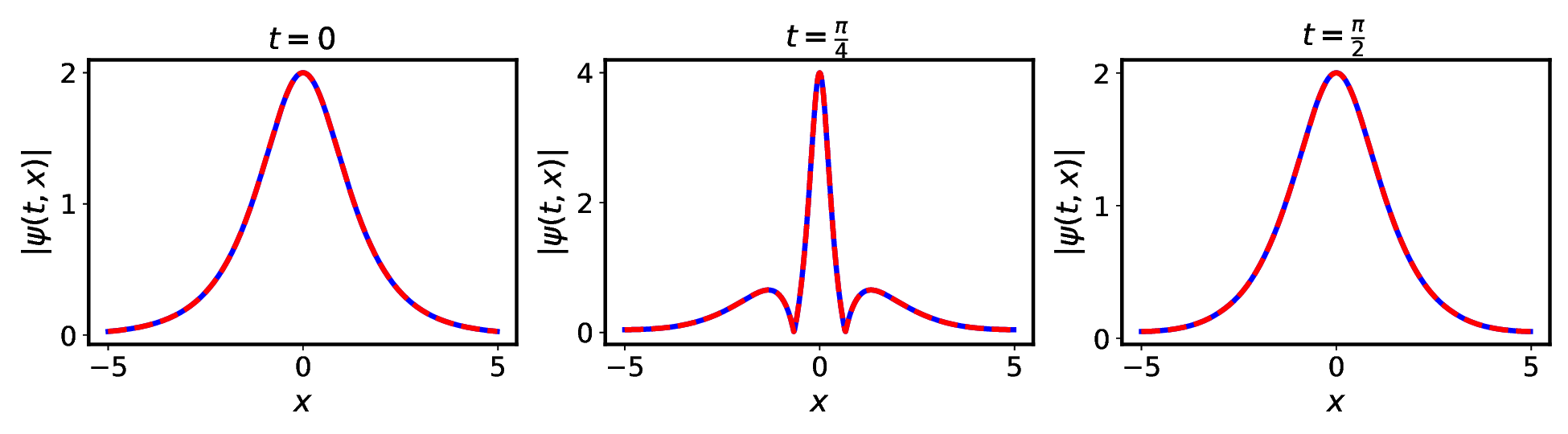}
$b$
\caption{(Color online) NLS equation: (a) The reference solution, predicted solution obtained using the Causal AS method and the absolute error; (b) Comparison of the predicted and reference solutions corresponding to the three temporal snapshots at $t=0,\frac{\pi}{4},\frac{\pi}{2}$.}
\label{fig3-7}
\end{figure}

\begin{figure}[htbp]
\centering
\includegraphics[width=6cm,height=4.5cm]{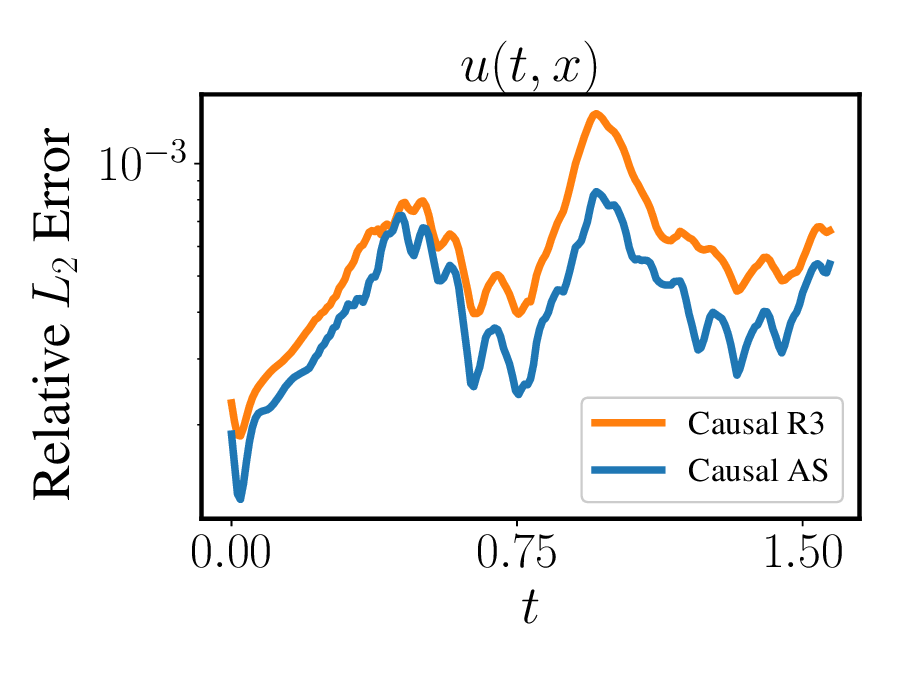}
$a$
\includegraphics[width=6cm,height=4.5cm]{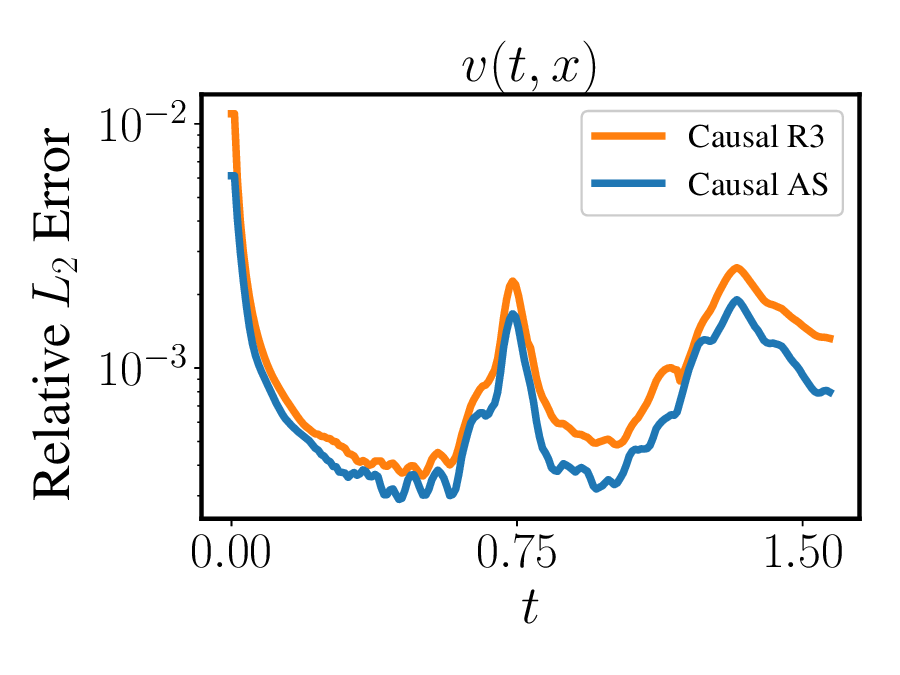}
$b$
\caption{(Color online) NLS equation: Relative $\mathbb{L}_2$ errors of (a) the real part $u(t,x)$ and (b) the imaginary part $v(t,x)$ obtained by Causal R3 and Causal AS ($N_t=20, N_x=250, N_A=1000$).}
\label{fig3-8}
\end{figure}

Initially, we employ the Causal AS method, configured with $N_A=1000$ and $p=1$. Fig. \ref{fig3-7} illustrates the corresponding experimental results. For a more comprehensive analysis, the Causal R3 method is also utilized while maintaining the same total number of sampling points, iterations, and other parameter settings. The errors in the real and imaginary parts of the predicted solutions at various time points for both methods are  depicted in Fig. \ref{fig3-8}, indicating that the Causal AS method achieves higher accuracy. 

Moreover, the Causal AS method with an updated $p$ starting from an initial value $p_0 = 1$ is adopted, with $\beta_1 = \beta_2 = 0.1$ and $\kappa = 0.9$. The number of adaptive collocation points $N_A$ ranges from 1000 to 4000 in intervals of 1000. The results of the various methods are summarized in Table \ref{table3-3}, where the values before and after the slash correspond to the errors of the real and imaginary parts, respectively. In short, the Causal AS method consistently exhibits superior accuracy compared to the Causal R3 method.

\begin{table}[htbp]
\caption{Relative $\mathbb{L}_2$ errors of the rogue wave solution for the NLS equation by Causal R3 and Causal AS.}
\label{table3-3} 
\centering
\begin{tabular}{cccccc}
\bottomrule
Method    & $N_A=1000$ & $N_A=2000$ & $N_A=3000$ & $N_A=4000$  \\ \hline
Causal R3 & 5.562e-04/7.895e-04     & 1.084e-03/1.875e-03     & 4.252e-04/5.986e-04     & 4.671e-04/6.771e-04          \\
Causal AS ($p\equiv 1$) &  3.928e-04/5.642e-04    & 3.603e-04/4.751e-04     & 3.747e-04/4.802e-04     & \pmb{4.041e-04/5.683e-04}         \\
Causal AS (Updating $p$) & \pmb{3.781e-04/5.299e-04}     & \pmb{3.558e-04/4.638e-04}     & \pmb{3.575e-04/4.645e-04}     & 4.326e-04/5.910e-04       \\ \toprule
\end{tabular}
\end{table}

\subsection{Korteweg-de Vries equation}\label{section_KdV}
\quad

The Korteweg-de Vries (KdV) equation \cite{KdV1895} is a nonlinear partial differential equation known for describing the propagation of shallow water waves by incorporating nonlinear and dispersive effects. One of its remarkable features is the presence of soliton solutions \cite{KdVsoliton1,KdVsoliton2}, a localized wave packet that can maintain its shape and speed over long periods. Additionally, the KdV equation is a completely integrable system, possessing an infinite number of conservation laws \cite{KdVconservation}. Here, the KdV equation takes the form \cite{KdVsoliton1}
\begin{align}
u_t+uu_x+\delta^2 u_{xxx}=0,	
\end{align}
with $\delta=0.022$, and satisfies the initial condition of $u(0, x)=\cos (\pi x)$ and the periodic boundary conditions.

\begin{figure}[htbp]
\centering
\includegraphics[width=16cm,height=4.2cm]{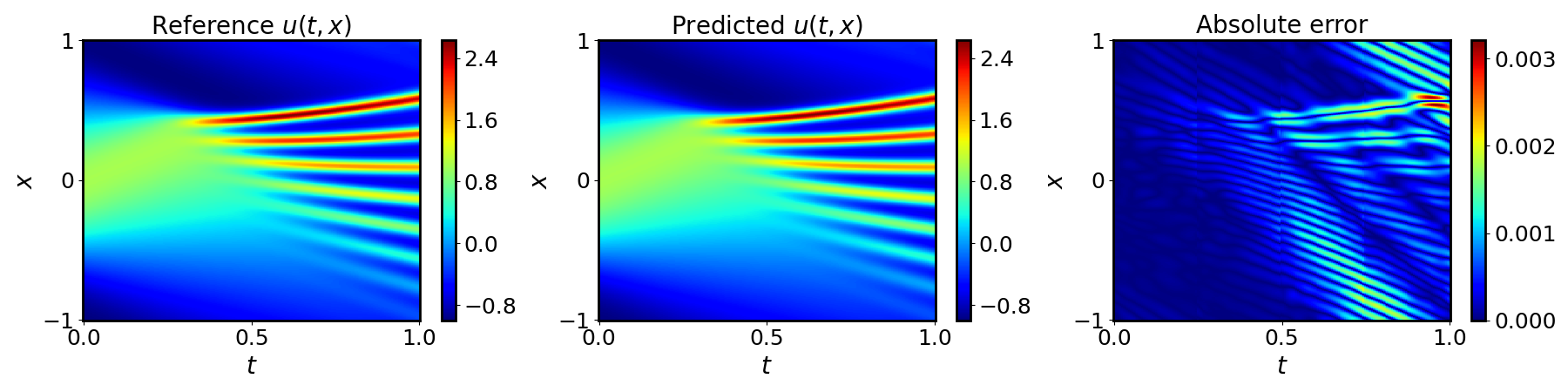}
$a$
\includegraphics[width=14cm,height=4cm]{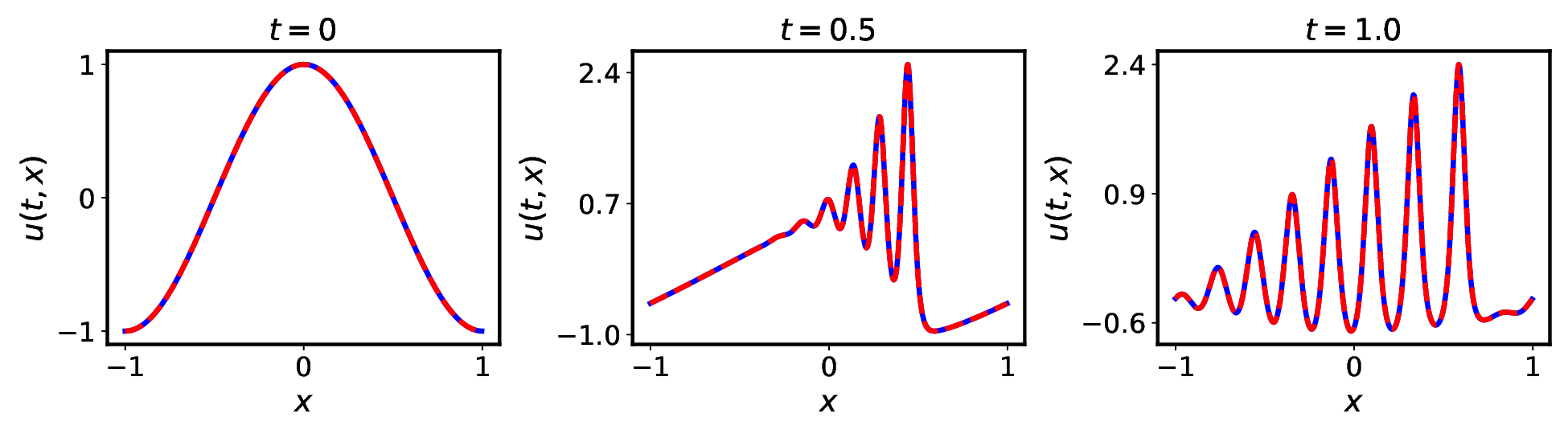}
$b$
\caption{(Color online) KdV equation: (a) The reference solution, predicted solution obtained using the Causal AS method and the absolute error; (b) Comparison of the predicted and reference solutions corresponding to the three temporal snapshots at $t=0,0.5,1.0$.}
\label{fig3-9}
\end{figure}

The reference solution for this equation is generated using the Fourier pseudo-spectral method. The spatial domain is $x \in [-1,1]$ with 512 Fourier modes, and the temporal domain is $t \in [0,1]$ with time-step $\Delta t = 0.005$. Then we employ the Causal AS method with $p$ fixed as a constant 1 to numerically solve the KdV equation. Analogous to Section \ref{section_NLS}, we also implement a time-marching strategy by dividing the time domain into four equal time windows, subsequently advancing the training sequentially in time. Furthermore, parameters such as the number of $\mathcal{T}_f$ and $N_{ic}$ as well as the architecture of neural networks remain consistent with the previous example and will not be elaborated upon further. During the network training process, an annealing strategy is utilized with an increasing sequence of $\{\epsilon_i\}_{i=1}^k = [10^{-2}, 10^{-1}, 10^0, 10^1, 10^2]$, which progressively strengthens the enforcement of the PDE residual constraint. The number of adaptive collocation points $N_A$ ranges from 1000 to 5000 in increments of 1000. Ultimately, the Causal AS method successfully captures the soliton solutions of the KdV equation, achieving relative $\mathbb{L}_2$  errors consistently on the order of $10^{-4}$. For $N_A = 1000$, the numerical results are depicted in Fig. \ref{fig3-9}, including the predicted solution, the absolute error, and the temporal evolution of the solution. The overall relative $\mathbb{L}_2$ error throughout the spatio-temporal domain is $7.455e-04$. We can observe the gradual transformation from the initial cosine wave into a train of eight solitons, which developed from the oscillations.

Maintaining consistency in all parameter settings, including but not limited to the total number of collocation points and the total iterations per subregion, we proceed to utilize the Causal R3 method to simulate the soliton solutions of the KdV equation. For $N_A = 1000$, the Causal R3 method achieves a relative $\mathbb{L}_2$ error of solely 2.053e-02 in prediction. The Causal R3 method maintains the same number of iterations per time window as the Causal AS method; however, the use of an annealing strategy in the Causal AS method results in differing iteration counts across subregions. Moreover, we note significant disparities in the iteration counts across the four subregions in this example, with a pronounced increasing trend rather than similar magnitudes. The neural network training for the first time window converges after approximately $7 \times 10^4$ iterations, while that for the final one demands up to $6 \times 10^5$ iterations. This could be attributed to the increasing difficulty of network training as the number of generated solitons rises. Nevertheless, the difference in the number of iterations among subregions  affects the accuracy of the Causal R3 method, particularly when the iteration counts for the first subregion are substantially lower. Insufficient training in the first subregion can lead to notable errors in the pseudo-initial values for the following subregions, thereby affecting the subsequent training results and causing the errors to accumulate progressively over time.

\begin{figure}[htbp]
\centering
\includegraphics[width=16cm,height=4.2cm]{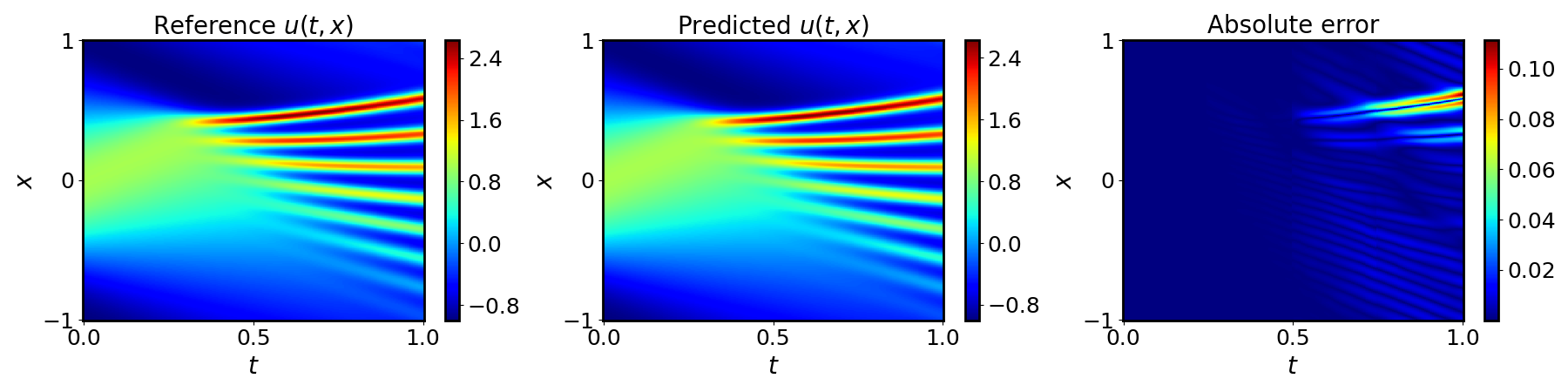}
\caption{(Color online) KdV equation: The reference solution, predicted solution obtained using the Causal R3 method and the absolute error.}
\label{fig3-10}
\end{figure}

To address this issue, we evenly allocate the total iteration counts across all sub-domains in the Causal AS method to each sub-domain of the Causal R3 method. The outcomes after this adjustment are depicted in Fig. \ref{fig3-10}. It is apparent that the prediction accuracy deteriorates near the final time, with the absolute error notably higher than that shown in Fig. \ref{fig3-9} (a). Although the relative $\mathbb{L}_2$ error has decreased to 1.256e-02—an improvement compared to the accuracy when the Causal R3 method matched the iteration counts per subregion with the Causal AS method—the overall accuracy is still notably worse than that achieved by the Causal AS method.

\begin{figure}[htbp]
\centering
\includegraphics[width=6cm,height=4.5cm]{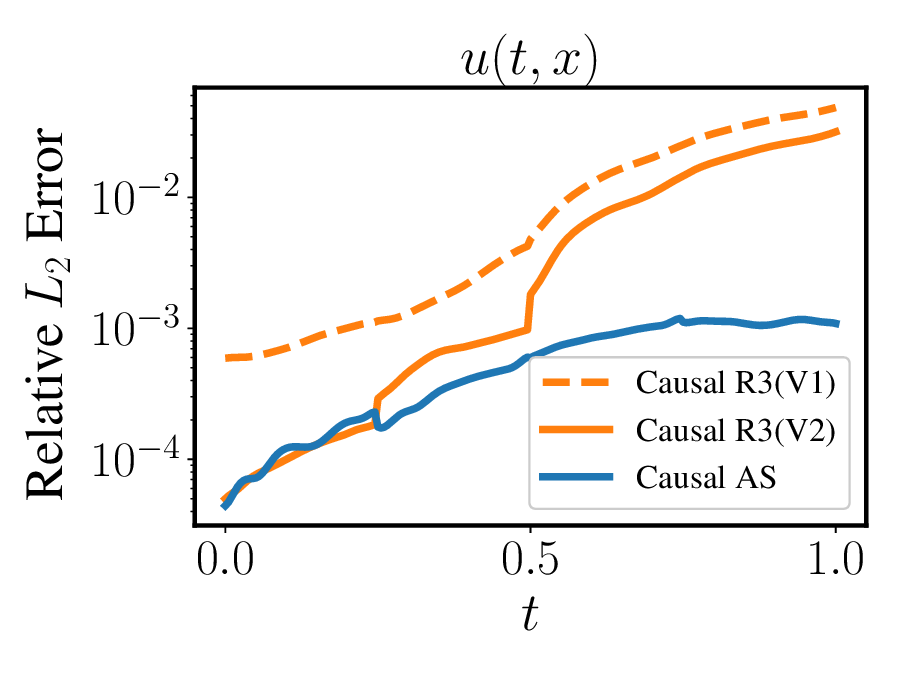}
$a$
\includegraphics[width=6cm,height=4.5cm]{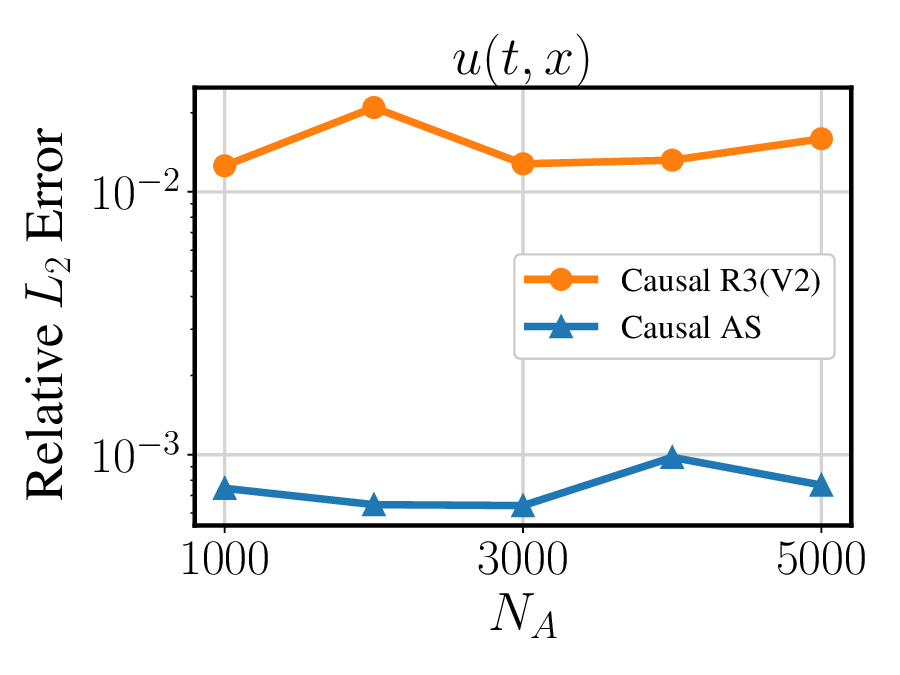}
$b$
\caption{(Color online) KdV equation: (a) Relative $\mathbb{L}_2$ errors of $u(t,x)$ obtained by Causal R3 and Causal AS ($N_t=20, N_x=250, N_A=1000$); (b) Relative $\mathbb{L}_2$ errors of $u(t,x)$ with different number of adaptive collocation points.}
\label{fig3-11}
\end{figure}

Fig. \ref{fig3-11}(a) depicts the errors at various time points of the aforementioned methods. The Causal R3 method with the same number of iterations per time window as the Causal AS method is denoted by Causal R3(V1), while that with equal iterations per sub-domain is denoted as Causal R3(V2). Obviously, for Causal R3(V2), the error in the first subregion is even slightly lower than that of the Causal AS method, owing to the increased number of iterations. However, the errors still exhibit a significant increase starting from the second subregion.

During the training process, a learning rate decay strategy is adopted to dynamically adjust the learning rate, promoting rapid convergence to an optimal solution in the early stages and stable convergence in the later stages. More precisely, we employ an exponential learning rate decay with an initial learning rate of 0.001, a decay rate of 0.9, and update the learning rate every 5000 steps. For Causal R3(V2), the neural network for each subregion performs approximately 420000 iterations, resulting in a learning rate of approximately $1.433 \times 10^{-7}$ by the end of training. The learning rate at this magnitude renders parameter updates negligible and additional iterations fail to produce significant changes in the network parameters. Consequently, increasing the number of iterations further does not contribute to substantial improvements in accuracy for the Causal R3 method.

Ultimately, Table 4 and Fig. \ref{fig3-11} (b) present the relative $\mathbb{L}_2$ errors of two methods under different numbers of adaptive collocation points. These results highlight the remarkable accuracy advantage of the proposed Causal AS method in this example, whereas the performance of the Causal R3 method is less satisfactory.

\begin{table}[htbp]
\caption{Relative $\mathbb{L}_2$ errors of the soliton solution for the KdV equation by Causal R3 (V2) and Causal AS.}
\label{table3-4} 
\centering
\begin{tabular}{cccccc}
\bottomrule
Method    & $N_A=1000$ & $N_A=2000$ & $N_A=3000$ & $N_A=4000$ & $N_A=5000$ \\ \hline
Causal R3 & 1.256e-02    & 2.092e-02     & 1.278e-02     & 1.319e-02     & 1.591e-02      \\
Causal AS & 7.455e-04     & 6.461e-04     & 6.407e-04     & 9.778e-04     & 7.667e-04     \\ \toprule
\end{tabular}
\end{table}

\subsection{Modified Korteweg-de Vries equation}\label{section_mKdV}
\quad

The modified Korteweg-de Vries (mKdV) equation is a variant of the KdV equation. As an extension and complement to the KdV equation, it is employed to describe various types of nonlinear wave phenomena. The KdV and mKdV equations are interconnected through the Miura transform \cite{Miura}, a nonlinear transformation that converts solutions of the mKdV equation into those of the KdV equation. This connection unveils the profound mathematical relationship between the two equations, providing essential tools for further research. Furthermore, the mKdV equation, along with the previously mentioned KdV and NLS equations, belongs to the AKNS (Ablowitz-Kaup-Newell-Segur) family, a group of nonlinear partial differential equations that are integrable through the inverse scattering transform (IST) method \cite{IST}.

Then we investigate the effectiveness of the Causal AS method in numerically solving the mKdV equation
\begin{align}
	u_t+6u^2u_x+u_{xxx}=0.
\end{align}
The exact breather solution has already been derived using an analytical method \cite{mKdVbreather}.
\begin{align} 
u(t, x)&=2 \partial_x\left[\arctan \left(\frac{\beta}{\alpha} \frac{\sin (\alpha(x+\delta t))}{\cosh (\beta(x+\gamma t))}\right)\right] \\ & =2 \beta \operatorname{sech}(\beta(x+\gamma t))\times\left[\frac{\cos (\alpha(x+\delta t))-(\beta / \alpha) \sin (\alpha(x+\delta t)) \tanh (\beta(x+\gamma t))}{1+(\beta / \alpha)^2 \sin ^2(\alpha(x+\delta t)) \operatorname{sech}^2(\beta(x+\gamma t))}\right],
\end{align}
with $\delta=\alpha^2-3\beta^2$, $\gamma=3\alpha^2-\beta^2$ and $\alpha, \beta \in \mathbb{R} \backslash\{0\}$.

We select the computational domain as $(t,x) \in [-0.3,0.3] \times [-5,5]$ and choose parameters $\alpha=1.5$ and $\beta=1$ to obtain the corresponding initial and boundary data. The time domain is divided into 4 equal segments, each with an interval length of 0.15. For each time window, we create a uniform mesh of size $N_t=20$ in the time domain and $N_x = 250$ in the spatial domain. Given the non-periodic boundary condition in this example, it cannot be strictly enforced as a hard constraint. Taking the first sub-domain $(t,x) \in [-0.3,-0.15] \times [-5,5]$ as an example, the loss function is defined as follows:
\begin{align}
\mathcal{L}(\boldsymbol{\theta})=\lambda_{i c} \mathcal{L}_{i c}(\boldsymbol{\theta})+\lambda_{b c} \mathcal{L}_{b c}(\boldsymbol{\theta})+\lambda_r \mathcal{L}_r(\boldsymbol{\theta}),
\end{align}
where
\begin{align}
\begin{gathered}
\mathcal{L}_{i c}(\boldsymbol{\theta})=\frac{1}{N_{i c}} \sum_{i=1}^{N_{i c}}\left( \left|u_{\boldsymbol{\theta}}\left(-0.3, x_{i c}^i\right)-u_{ic}^i \right|^2 \right),\\	
\mathcal{L}_{b c}(\boldsymbol{\theta})=\frac{1}{2\cdot N_t}\sum_{i=1}^{N_t} w_i \left( \mathcal{L}_{b c}(t_i, -5, \boldsymbol{\theta})+\mathcal{L}_{b c}(t_i, 5, \boldsymbol{\theta}) \right),\\
\mathcal{L}_{r}(\boldsymbol{\theta})	= \frac{1}{N_x\cdot N_t +N_A} \sum_{i=1}^{N_t} \sum_{j=1}^{|\tau_i|} w_i \mathcal{L}_r\left(t_i, x_{i,j}, \boldsymbol{\theta}\right),\\
w_i=\exp \left(-\epsilon	 \sum_{k=1}^{i-1}  \left(\frac{1}{N_x} \sum_{j=1}^{|\tau_k|} \mathcal{L}_r\left(t_k, x_{k,j}, \boldsymbol{\theta}\right)+ \frac{1}{2} \sum_{x \in \{-5,5\}} \mathcal{L}_{b c} \left( t_k, x, \boldsymbol{\theta} \right) \right)   \right),
\end{gathered}
\end{align}
with $\mathcal{L}_{b c} \left( t_i, x_{bc}, \boldsymbol{\theta} \right)=\left| u_{\boldsymbol{\theta}}(t_i, x_{b c})-u_{bc}^i \right|^2, x_{bc} \in \{-5,5 \}$. Here, the parameters are set as $N_{ic}=256$, $\lambda_{ic}=100\), and \(\lambda_{bc}=\lambda_{r}=1$. We use a fully-connected neural network $u_{\boldsymbol{\theta}}$ with the $\tanh$ activation function to represent the solution. This network consists of 4 hidden layers with 128 neurons per layer.

\begin{figure}[htbp]
\centering
\includegraphics[width=16cm,height=4.2cm]{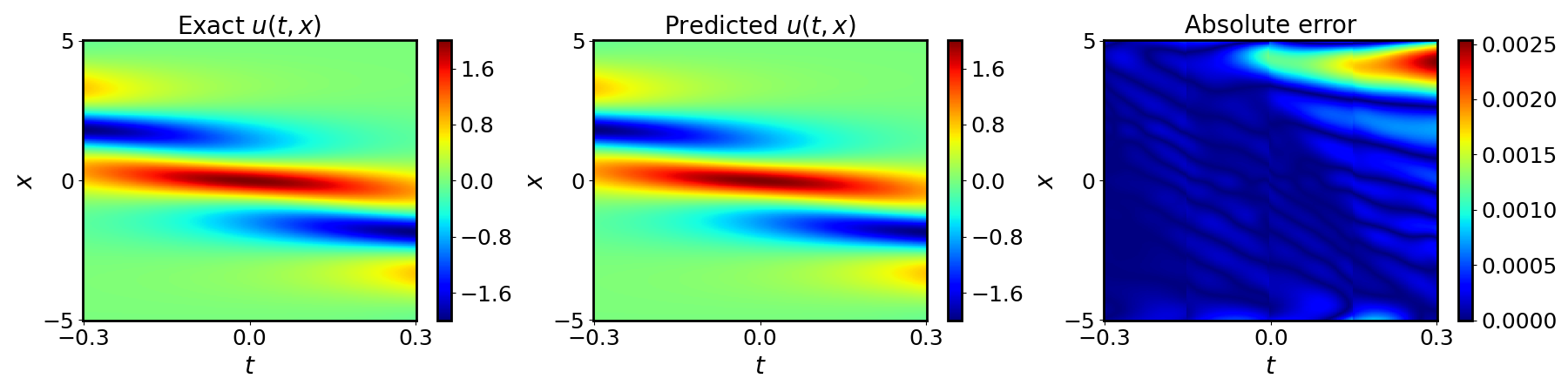}
$a$
\includegraphics[width=14cm,height=4cm]{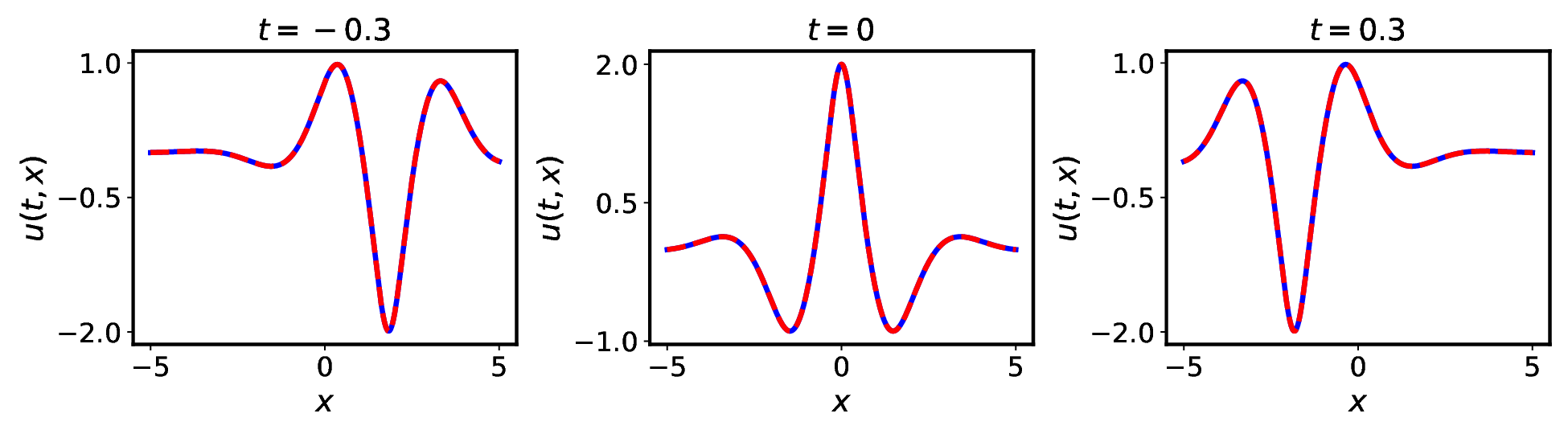}
$b$
\caption{(Color online) mKdV equation: (a) The exact solution, predicted solution obtained using the Causal AS method and the absolute error; (b) Comparison of the predicted and exact solutions corresponding to the three temporal snapshots at $t=-0.3,0,0.3$.}
\label{fig3-12}
\end{figure}

After setting $K=500, \rho=\frac{1}{4}$, the Causal AS method with $p=1$ is utilized to simulate the breather solution of the mKdV equation, with the number of adaptive collocation points $N_A$ ranging from 1000 to 5000, in increments of 1000. Fig. \ref{fig3-12} shows the experimental results of Causal AS with $N_A=1000$. It is evident that the predicted solution achieves an excellent agreement with the exact solution, resulting in an approximation error of $7.073e-04$.

\begin{figure}[htbp]
\centering
\includegraphics[width=16cm,height=4.2cm]{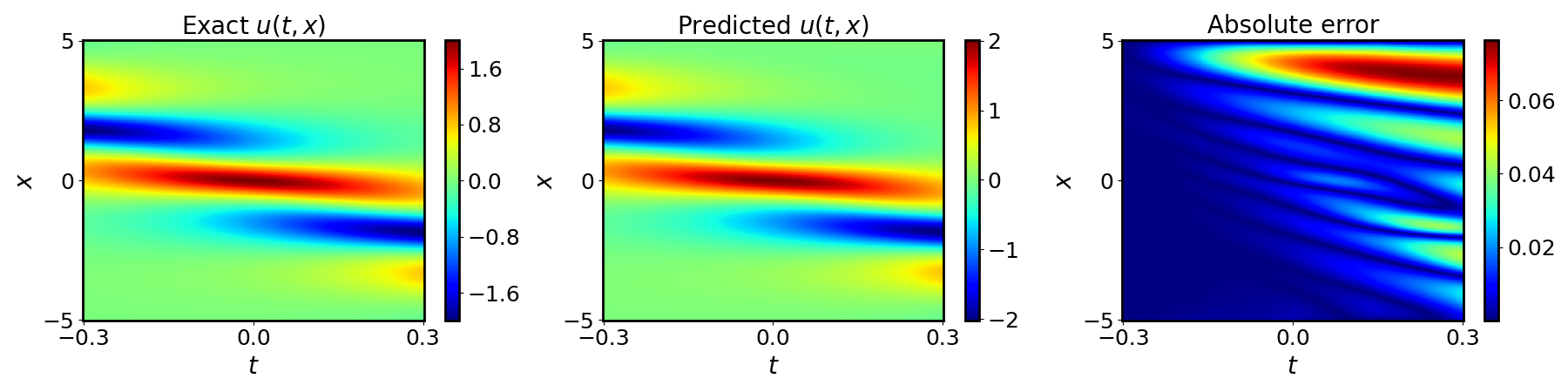}
$a$
\includegraphics[width=14cm,height=4cm]{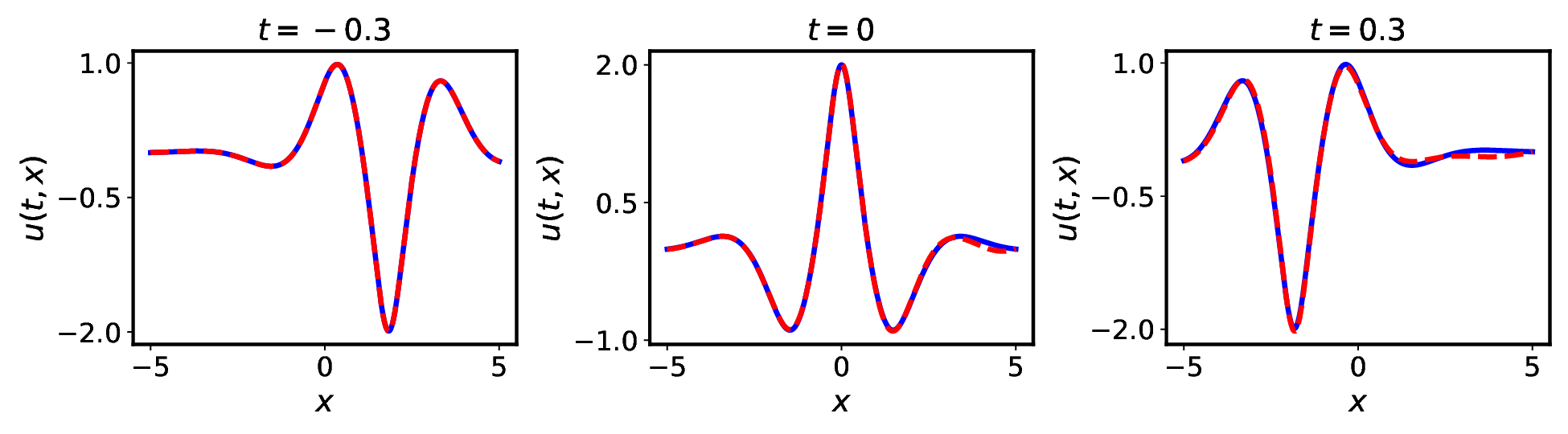}
$b$
\caption{(Color online) mKdV equation: (a) The exact solution, predicted solution obtained using the Causal R3 method and the absolute error; (b) Comparison of the predicted and exact solutions corresponding to the three temporal snapshots at $t=-0.3,0,0.3$.}
\label{fig3-13}
\end{figure}

\begin{figure}[htbp]
\centering
\includegraphics[width=6cm,height=4.5cm]{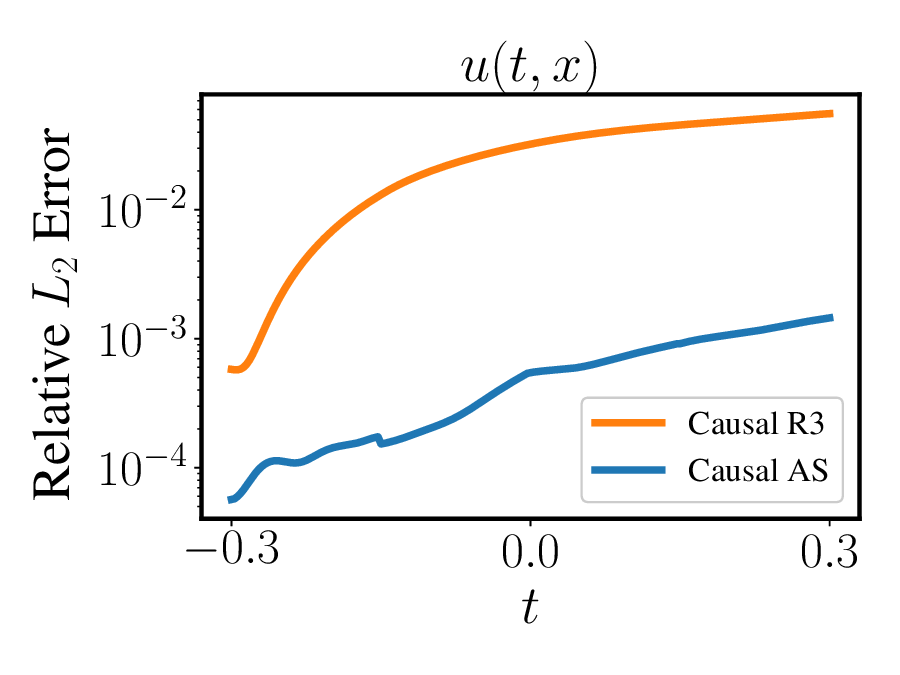}
$a$
\includegraphics[width=6cm,height=4.5cm]{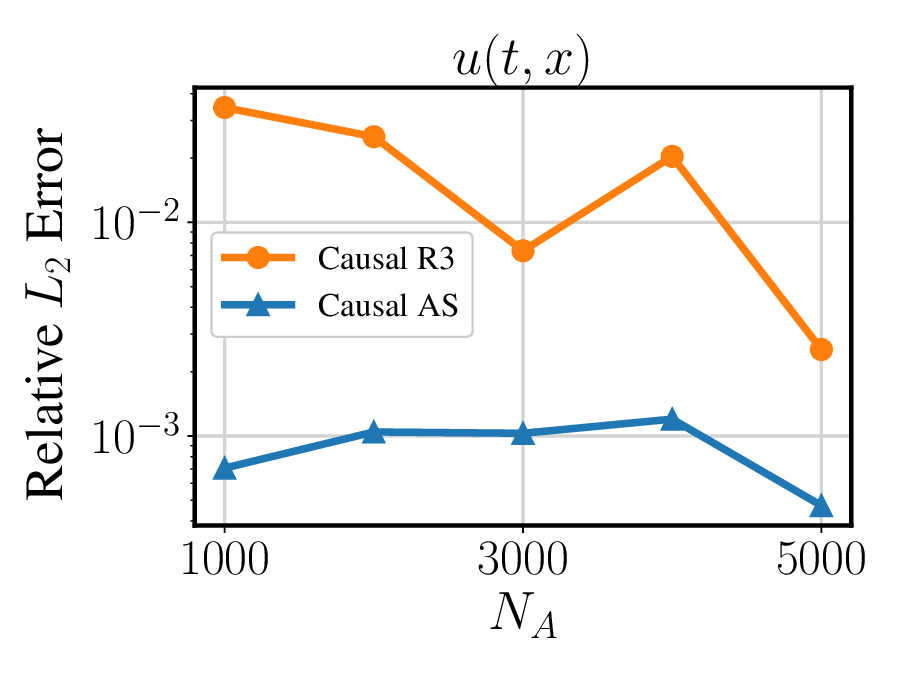}
$b$
\caption{(Color online) mKdV equation: (a) Relative $\mathbb{L}_2$ errors of $u(t,x)$ obtained by Causal R3 and Causal AS ($N_t=20, N_x=250, N_A=1000$); (b) Relative $\mathbb{L}_2$ errors of $u(t,x)$ with different number of adaptive sampling points.}
\label{fig3-14}
\end{figure}

The Causal R3 method is additionally applied to numerically solve the mKdV equation under identical parameter settings. In Fig. \ref{fig3-13}, we present the predicted solution with $N_A$ taken as 1000, as well as the absolute error and the comparison between the predicted and exact solutions at three different temporal snapshots, yielding a relative $\mathbb{L}_2$ error of $3.446 \times 10^{-2}$. It becomes apparent that the alignment between the predicted solution and the exact solution by the Causal R3 method deteriorates as time progresses. A comparison of Fig. \ref{fig3-12} and Fig. \ref{fig3-13} reveals that both methods exhibit larger errors in regions with greater spatiotemporal coordinates. Nevertheless, the absolute error of the Causal AS method is negligible compared to that of the Causal R3 method. Fig. \ref{fig3-14} (a) further illustrates the relative $\mathbb{L}_2$  errors of both methods over time, revealing a pronounced gap in accuracy. We also provide a comparative summary of errors for different $N_A$ values in Table \ref{table3-5} and Fig. \ref{fig3-14} (b). These results underscore the remarkable effectiveness of the proposed Causal AS method, demonstrating its ability to accurately capture the complex dynamical behavior of the breather solution.

\begin{table}[htbp]
\caption{Relative $\mathbb{L}_2$ errors of the breather solution for the mKdV equation by Causal R3 and Causal AS.}
\label{table3-5} 
\centering
\begin{tabular}{cccccc}
\bottomrule
Method    & $N_A=1000$ & $N_A=2000$ & $N_A=3000$ & $N_A=4000$ & $N_A=5000$ \\ \hline
Causal R3 &  3.446e-02    & 2.516e-02     & 7.358e-03     & 2.035e-02     & 2.538e-03     \\
Causal AS & 7.073e-04     & 1.044e-03     & 1.029e-03     & 1.199e-03     & 4.721e-04     \\ \toprule
\end{tabular}
\end{table}

\section{Conclusion}\label{Conclusion}

Inspired by causal training, we put forth the causality-guided adaptive sampling (Causal AS) method for physics-informed neural networks in this paper. By analyzing the characteristics of three typical regions during the Causal PINN training process, we aim to focus more attention on areas with larger PDE residuals within the regions being trained and optimized. Therefore, the fundamental idea is to use the weighted residual as the indicator for the selection of the collocation points. In particular, we propose two approaches to determine the hyper-parameter $p$ involved: (i) setting it as a constant coefficient, and (ii) updating it based on the temporal alignment driven update (TADU) scheme. Furthermore, the original Causal PINN framework is extended to solve problems with non-periodic boundary conditions after the re-formulation of the loss function and temporal weights. The proposed Causal AS algorithm was employed to solve the forward problems of four different PDEs, including the Allen-Cahn equation, for which the original PINN method fails, as well as the well-known integrable equations: the KdV equation, mKdV equation, and NLS equation. A comparison of the errors between this method and the Causal R3 method highlighted the superior accuracy of the Causal AS method. Additionally, we showcased the distribution of newly added collocation points at different iterations of the training to illustrate that the selection of these points aligns with our intended objectives.

Building on the strategy proposed here, where weighted PDE residuals guide the selection of adaptive collocation points, one could develop a variant that leverage probability density functions for adaptive sampling. Such an approach would offer a better balance between local and global sampling demands, potentially enhancing both flexibility and robustness. Taking computational costs into account, this paper employs a basic MLP architecture. A more powerful structure, the modified MLP \cite{modifiedMLP}, can yield significantly higher accuracy and can be employed in subsequent studies. The numerical examples in this study are confined to 1+1 dimensional equations. In the future research, we will apply the adaptive sampling method proposed here to higher-dimensional PDEs to further investigate its effectiveness. Given the superior accuracy of the Causal PINN method in solving PDEs, several enhanced algorithms have emerged based on this framework, such as those discussed in \cite{improvedCausal}. The Causal AS strategy can also be adopted to these variants, potentially improving their performance.

\section*{Acknowledgments}
The project is supported by the National Natural Science Foundation of China (No. 12175069 and No. 12235007), Science and Technology Commission of Shanghai Municipality, China (No. 21JC1402500 and No. 22DZ2229014), and Natural Science Foundation of Shanghai, China (No. 23ZR1418100).

\appendix
\section{Symbol Description}

Table \ref{tableA-1} provides an overview of the main symbols and notations utilized in this study.

\begin{table}[htbp]
\caption{Nomenclature: Summary of the main symbols and notations used in this work.}
\label{tableA-1} 
\centering
\begin{tabular}{ll}
\bottomrule
Notation & Description \\ \hline
$w_i$        & The residual weights at time $t_i$       \\
$N_t$        & The number of temporal collocation points         \\
$N_x$        & The number of spatial collocation points         \\
$N_A$        & The number of adaptive collocation points         \\
$N_f$        & The total number of collocation points        \\
$\mathcal{T}_f$        & The initial set of collocation points         \\
$\mathcal{T}_A$        & The set of adaptive collocation points         \\
$\mathcal{T}_{total}$        & The set of all collocation points         \\
$\tau_i$        & The set of collocation points at time $t_i$         \\
$U$        & The randomly chosen dense set of adaptive collocation points        \\
$\rho$        & The proportional relationship between the number of points in sets $\mathcal{T}_A$ and $U$         \\
$p$        & The parameter controlling the dominant effect between weights and PDE residuals         \\
$\beta_1$        & The parameter of the maximum change factor for $p$         \\ 
$\beta_2$        & The parameter controlling the difficulty of achieving the maximum change factor for $p$         \\ 
$t_{ada}$        & The approximate time during which the model is being trained         \\ 
$t_{w}$        & The concentration time of the previous batch of adaptive collocation points        \\
$\kappa$        & The threshold of         \\
$\boldsymbol{\theta}$        & All trainable parameters of a neural network         \\ 
$\epsilon$        & Causality parameter         \\ 
$\delta$        & Stopping criterion threshold for terminating a training loop         \\\toprule
\end{tabular}
\end{table}

\section{Model settings} 

Table \ref{tableA-2} summarizes the boundary conditions for each example, as well as whether the annealing strategy is applied and the number of time windows. The hyper-parameters of the neural networks are listed in Table \ref{tableA-3}, where $N_t$ and $N_{bc}$ refer to the number of temporal collocation points and boundary points within each time window respectively, and "Max Iteration" is the maximum iteration for every tolerance $\epsilon$ in each time window.

\begin{table}[htbp]
\caption{Condition setup and strategy usage}
\label{tableA-2} 
\centering
\begin{tabular}{cccc}
\bottomrule
Case       & Boundary conditions & Annealing strategy & \# Time windows \\ \hline
Allen-Cahn & Periodic            & No                 & 1               \\ \hline
NLS        & Periodic            & Yes                & 4               \\ \hline
KdV        & Periodic            & Yes                & 4               \\ \hline
mKdV       & Non-periodic        & Yes                & 4               \\ \toprule
\end{tabular}
\end{table}

\begin{table}[htbp]
\caption{The network hyper-parameters}
\label{tableA-3} 
\centering
\begin{tabular}{cccccccc}
\bottomrule
Case       & Hidden layer & Neurons per layer & $N_t$ & $N_x$ & $N_{ic}$ & $N_{bc}$ & Max Iterations \\ \hline
Allen-Cahn &   4       & 128    & 100   & 256   & 256    & —    &$3 \times 10^5$                \\ \hline
NLS        & 3    &   128    &  20  &  250  &  256   &  —   &$3 \times 10^5$                \\ \hline
KdV        &  3     &  128    &  20  & 250   &  256   &  —   & $3 \times 10^5$               \\ \hline
mKdV       &  4      &   128     &  20  &  250  & 256    & 40    & $3 \times 10^5$               \\ \toprule
\end{tabular}
\end{table}

\section{Supplementary training results} 

\subsection{NLS equation}
\quad

\begin{figure}[htbp]
\centering
\includegraphics[width=7cm,height=4.5cm]{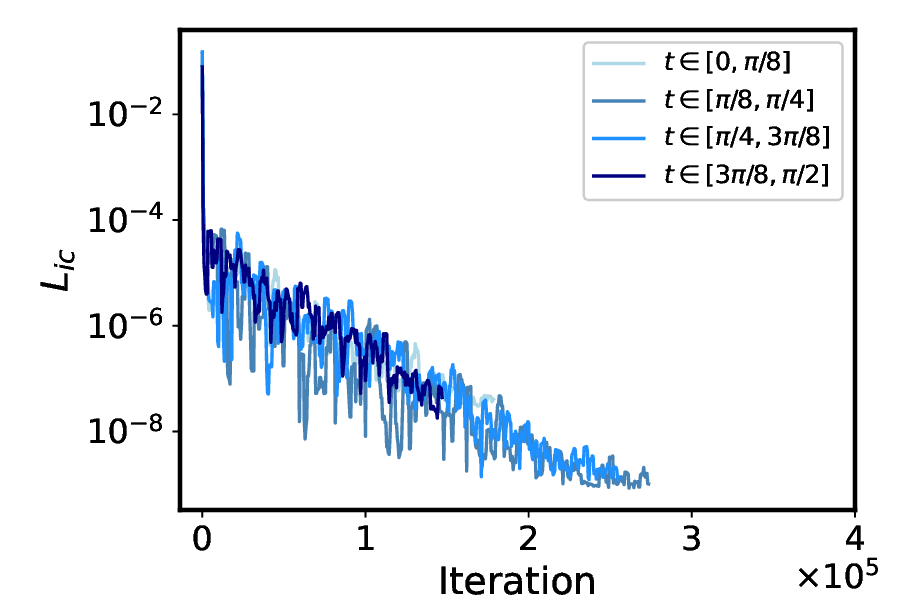}
\includegraphics[width=7cm,height=4.5cm]{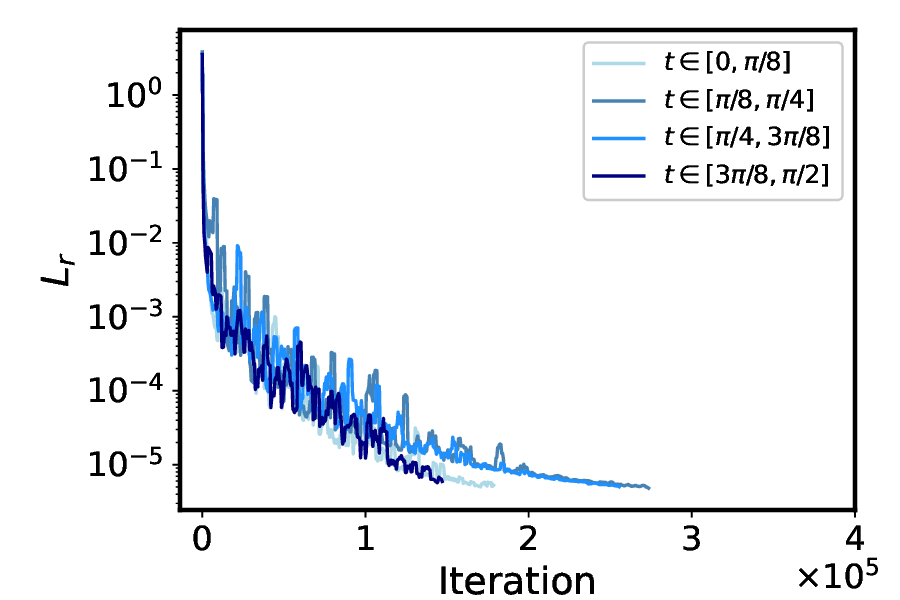}
$a$
\includegraphics[width=7cm,height=4.5cm]{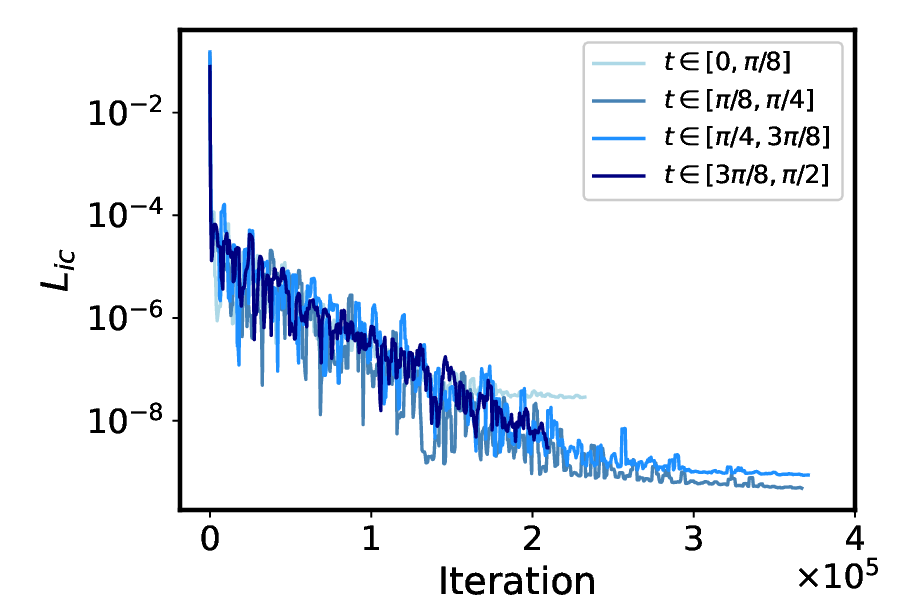}
\includegraphics[width=7cm,height=4.5cm]{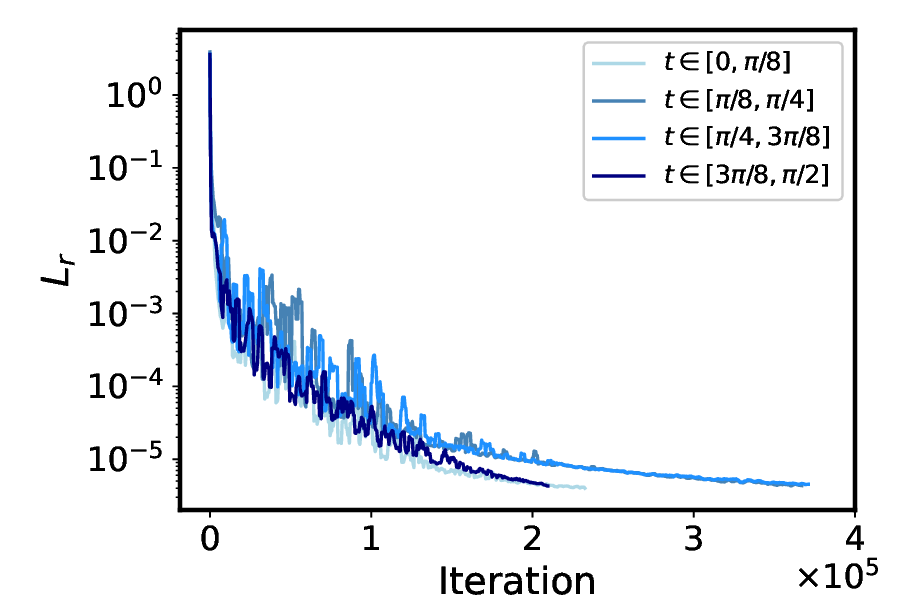}
$b$
\includegraphics[width=7cm,height=4.5cm]{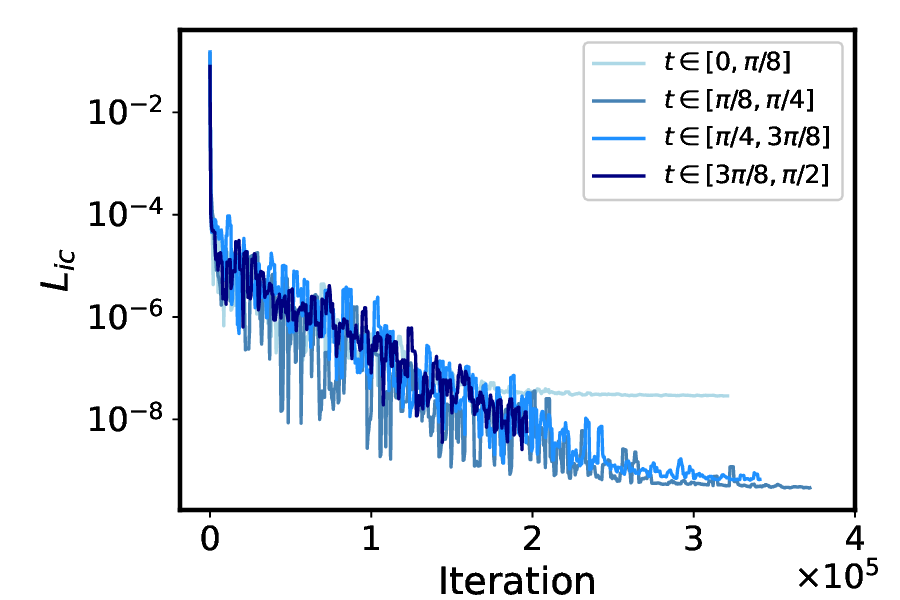}
\includegraphics[width=7cm,height=4.5cm]{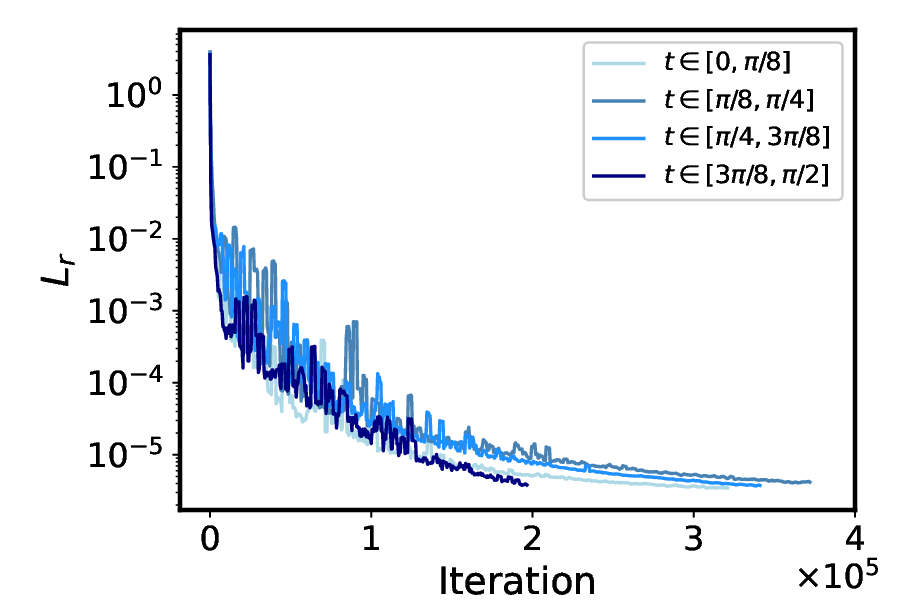}
$c$
\includegraphics[width=7cm,height=4.5cm]{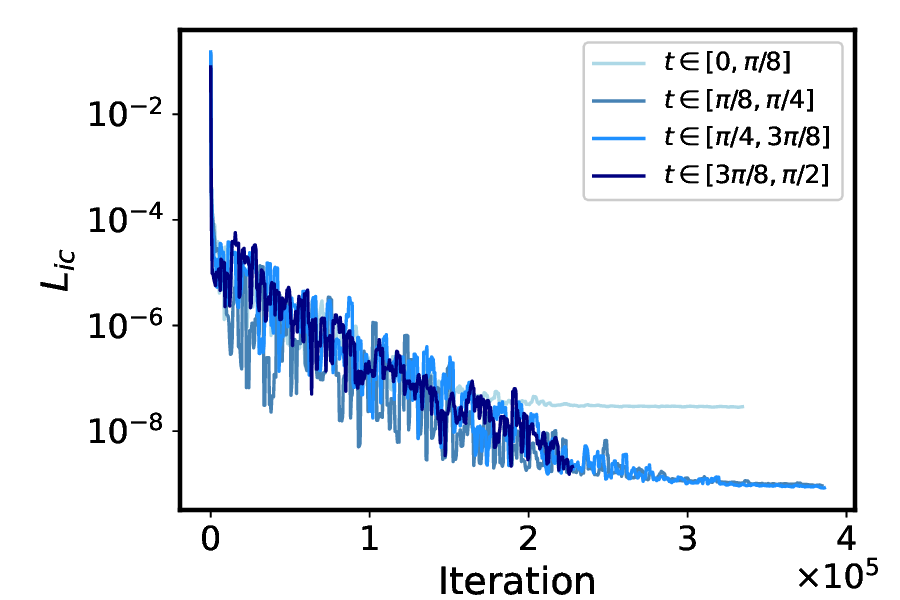}
\includegraphics[width=7cm,height=4.5cm]{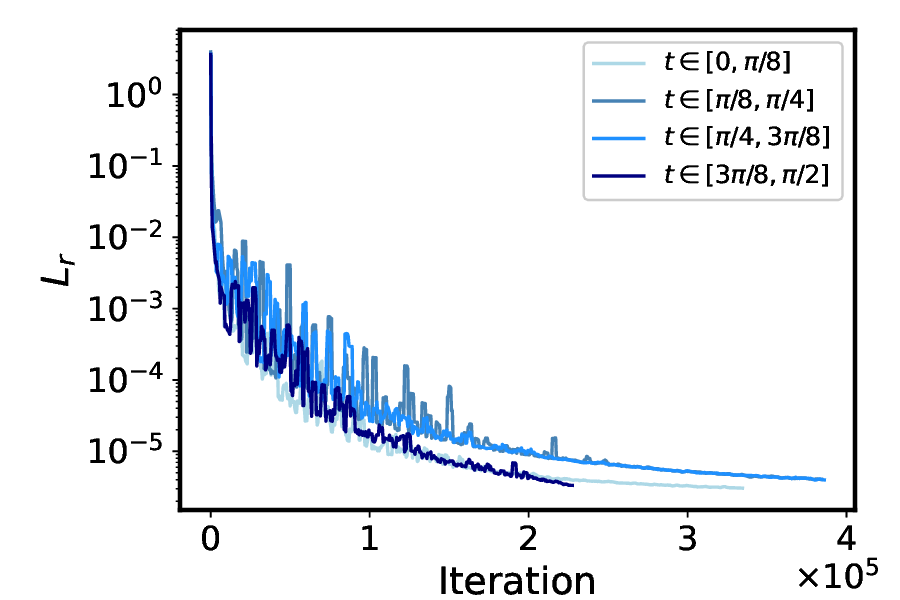}
$d$
\caption{(Color online) Loss convergence of Causal AS method with $p=1$ for NLS equation: (a) $N_A=1000$; (b) $N_A=2000$; (c) $N_A=3000$; (d) $N_A=4000$.}
\label{figA-1}
\end{figure}

Fig. \ref{figA-1} shows the evolution curves of the loss terms across different time windows when numerically solving the NLS equation with the Causal AS method.

\subsection{KdV equation}
\quad

\begin{figure}[htbp]
\centering
\includegraphics[width=7cm,height=4.5cm]{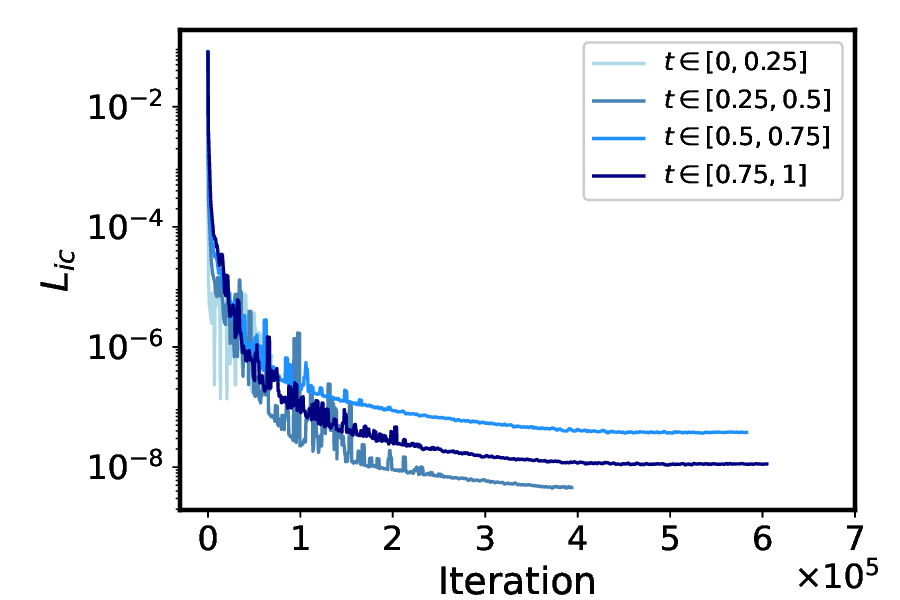}
\includegraphics[width=7cm,height=4.5cm]{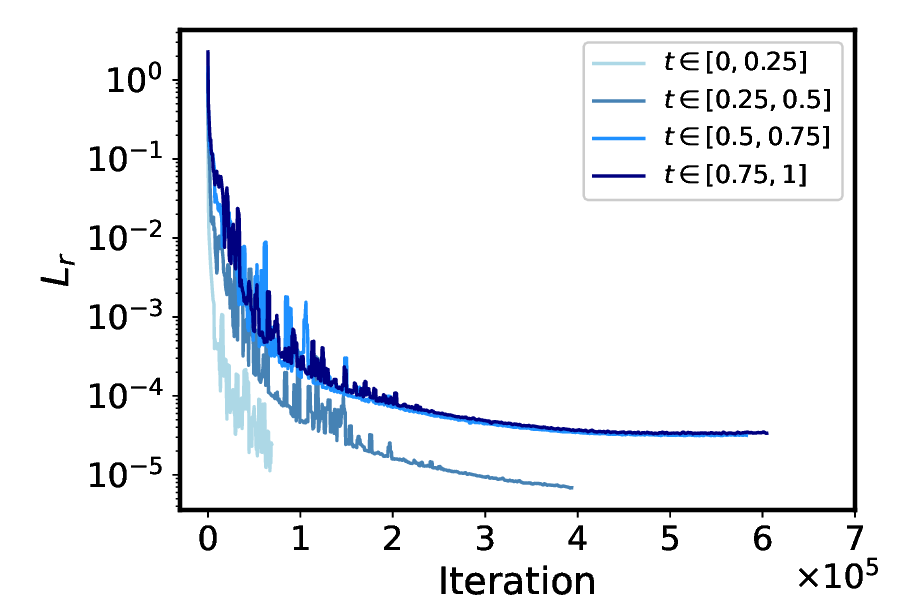}
$a$
\includegraphics[width=7cm,height=4.5cm]{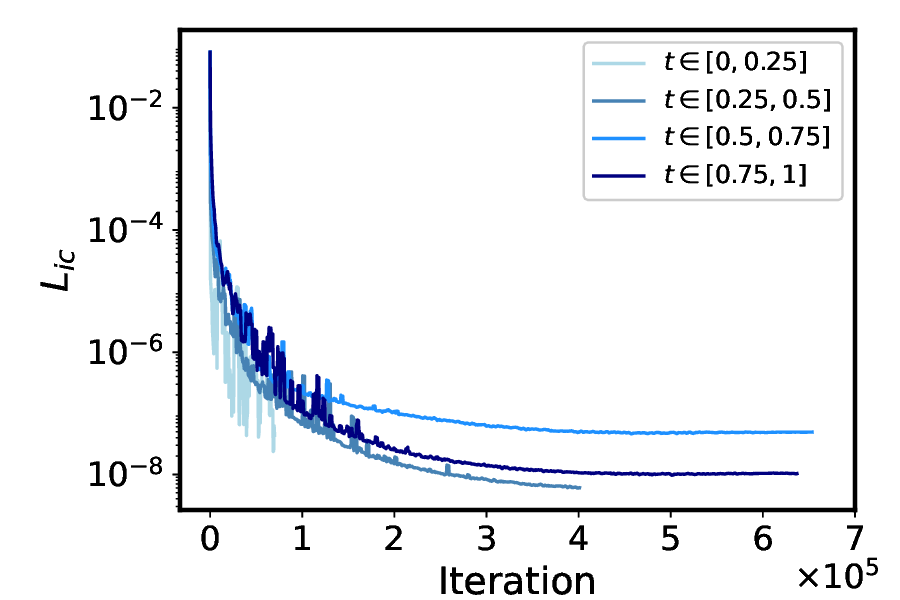}
\includegraphics[width=7cm,height=4.5cm]{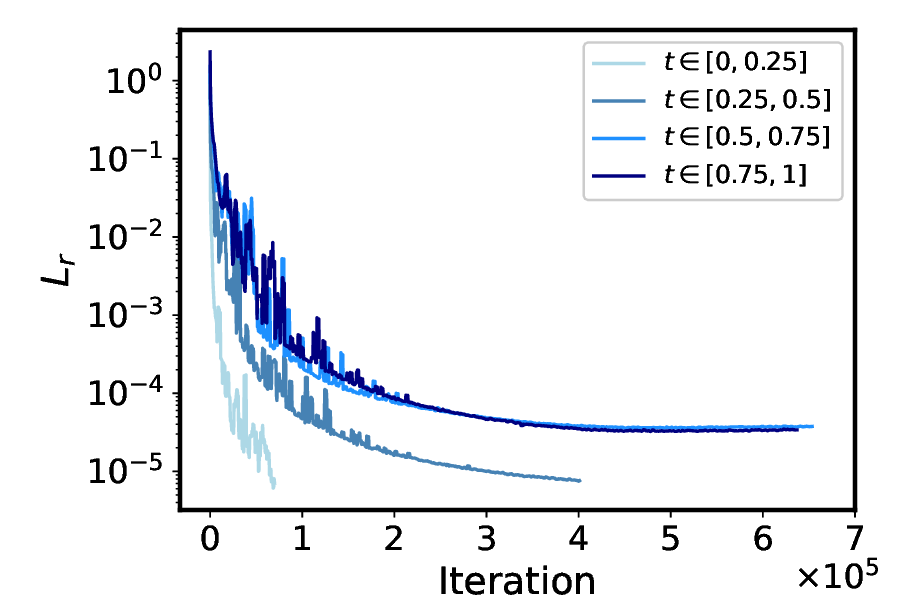}
$b$
\includegraphics[width=7cm,height=4.5cm]{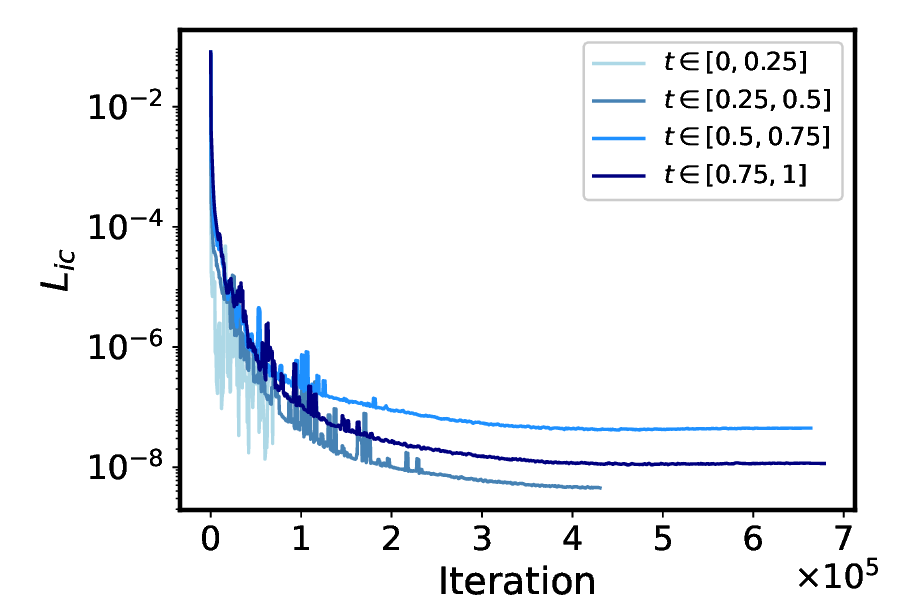}
\includegraphics[width=7cm,height=4.5cm]{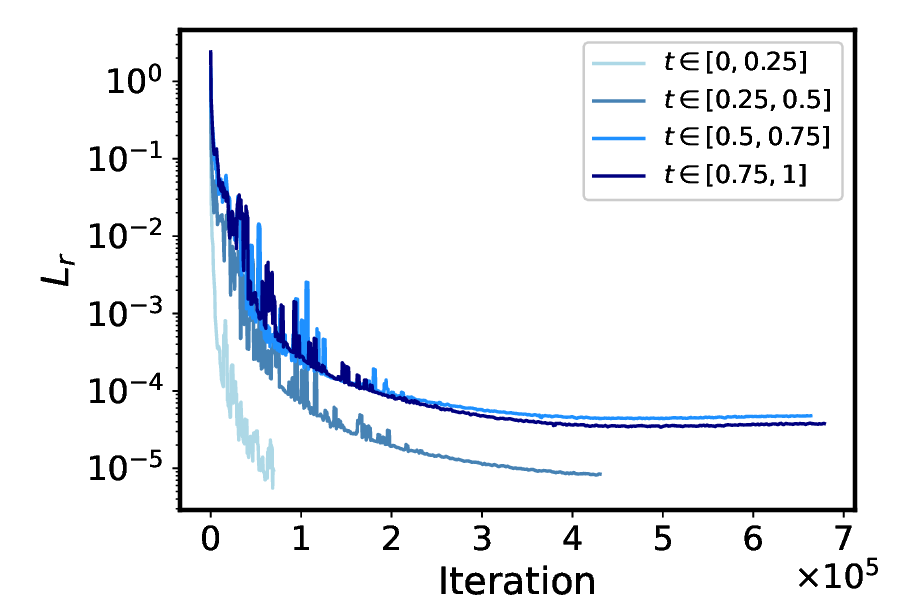}
$c$
\includegraphics[width=7cm,height=4.5cm]{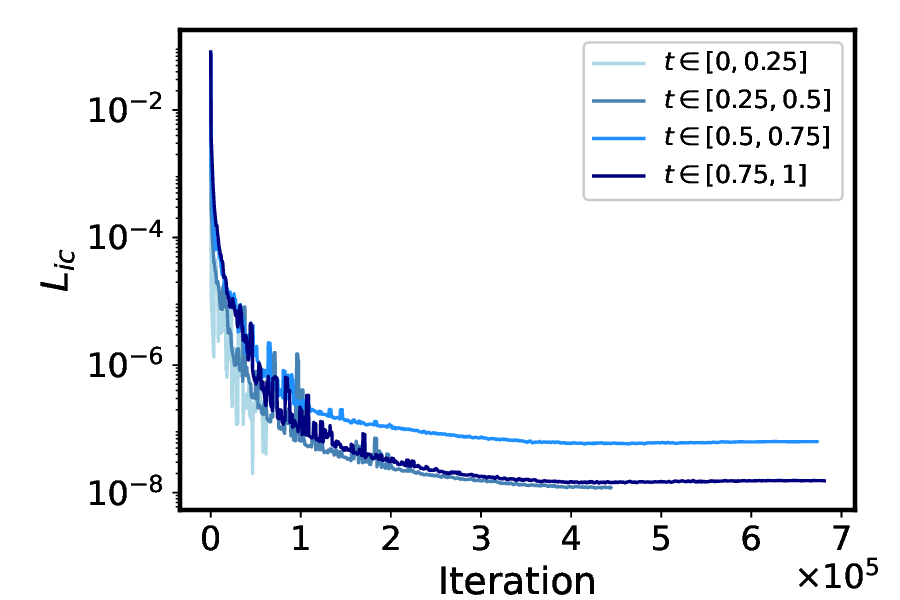}
\includegraphics[width=7cm,height=4.5cm]{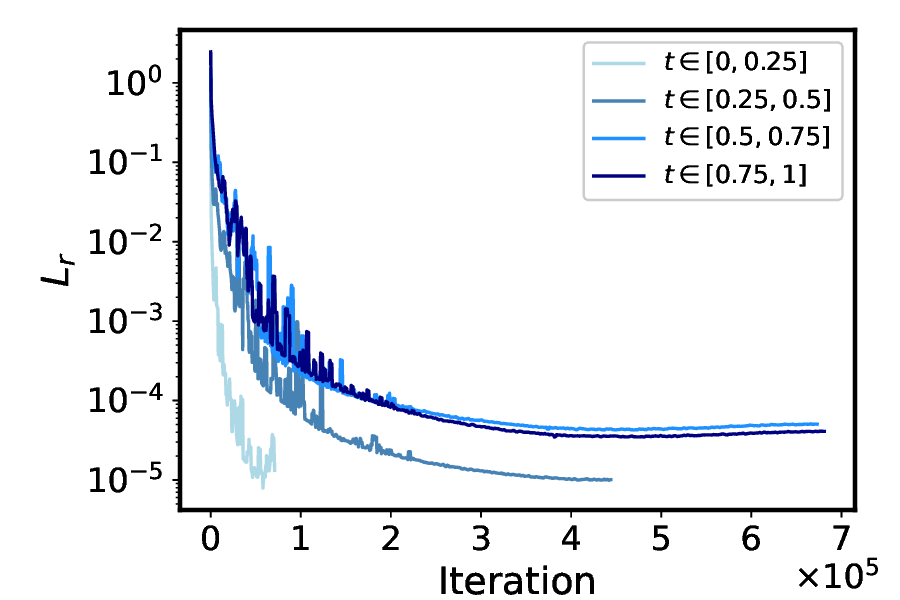}
$d$
\includegraphics[width=7cm,height=4.5cm]{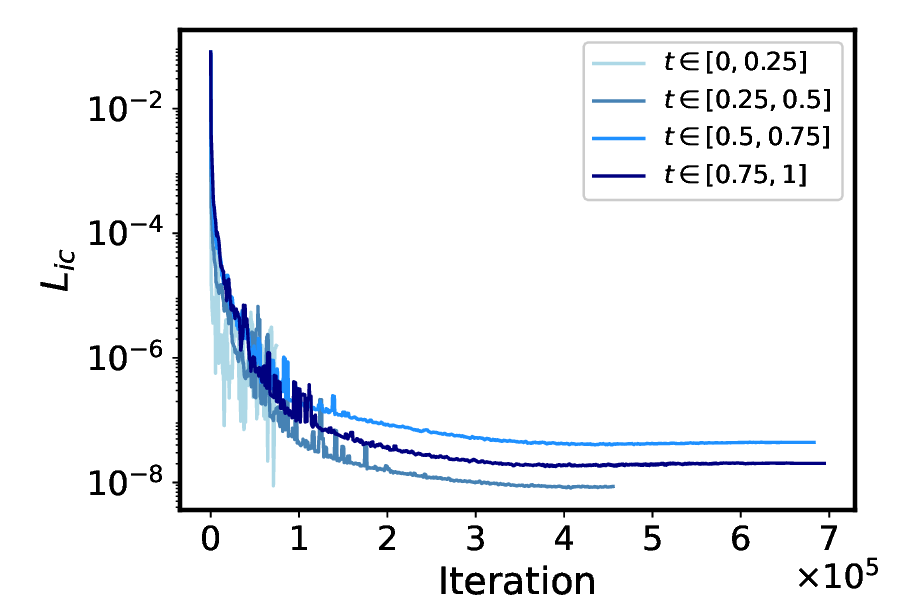}
\includegraphics[width=7cm,height=4.5cm]{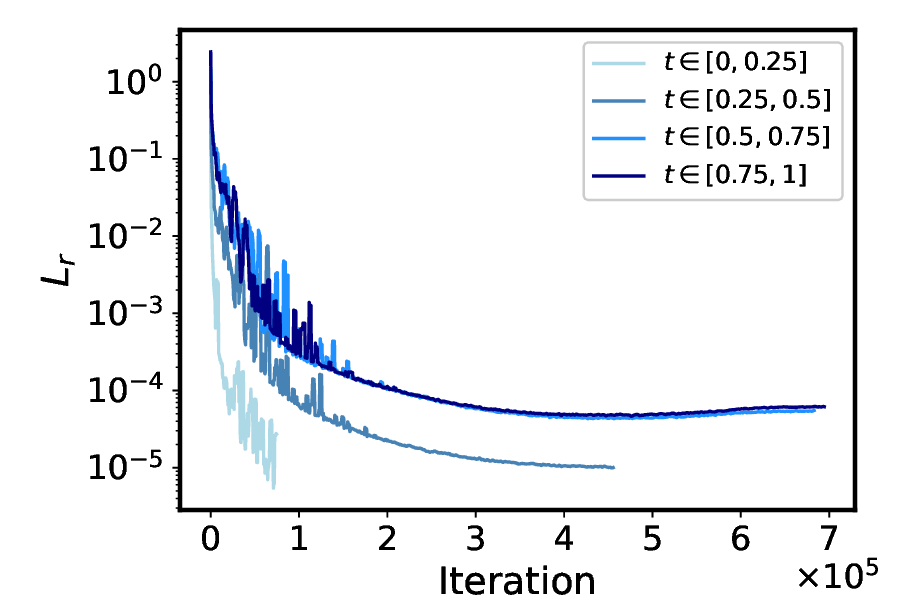}
$e$
\caption{(Color online) Loss convergence of Causal AS method with $p=1$ for KdV equation: (a) $N_A=1000$; (b) $N_A=2000$; (c) $N_A=3000$; (d) $N_A=4000$; (e) $N_A=5000$.}
\label{figA-2}
\end{figure}

Fig. \ref{figA-2} shows the evolution curves of the loss terms across different time windows when numerically solving the KdV equation with the Causal AS method.

\subsection{Modified KdV equation}
\quad

\begin{figure}[htbp]
\centering
\includegraphics[width=5.5cm,height=3.6cm]{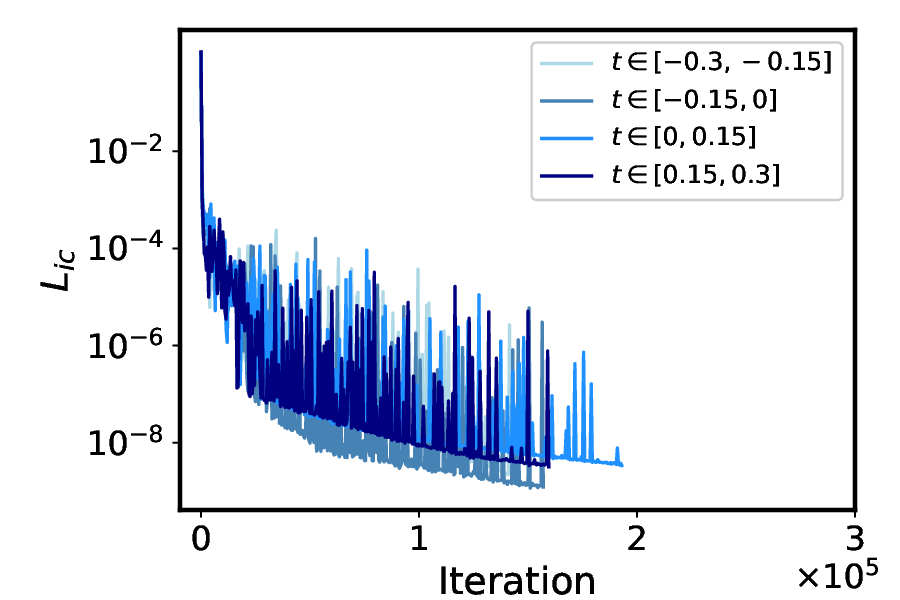}
\includegraphics[width=5.5cm,height=3.6cm]{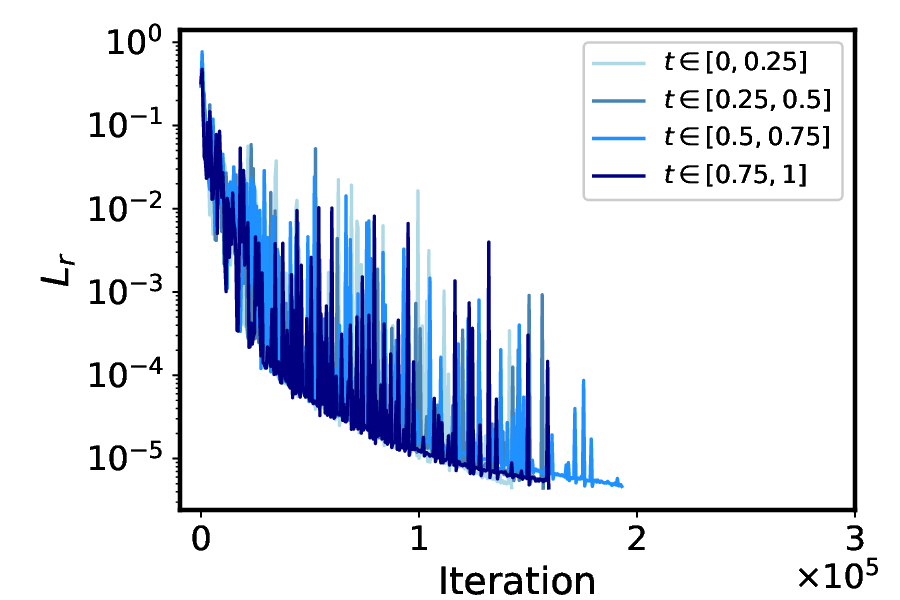}
\includegraphics[width=5.5cm,height=3.6cm]{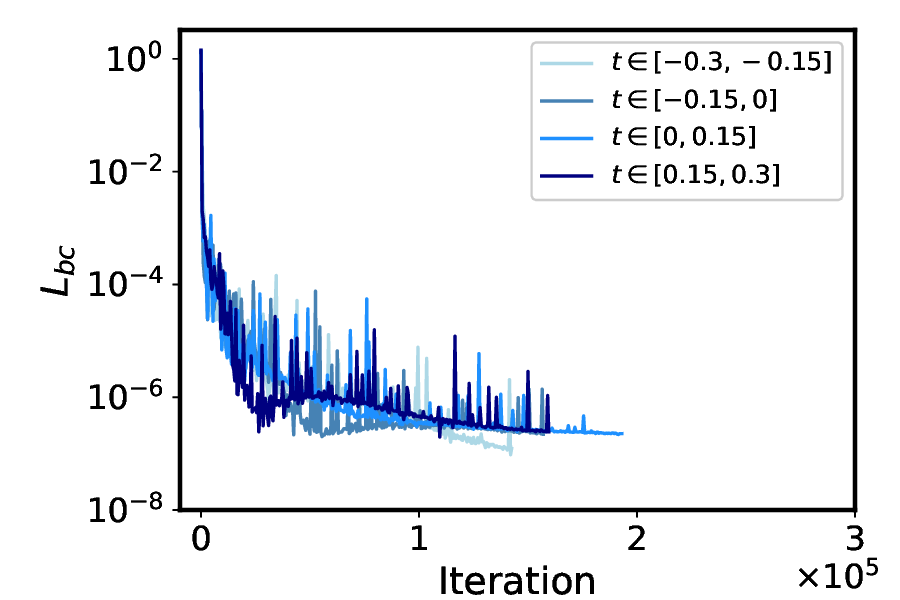}
$a$
\includegraphics[width=5.5cm,height=3.6cm]{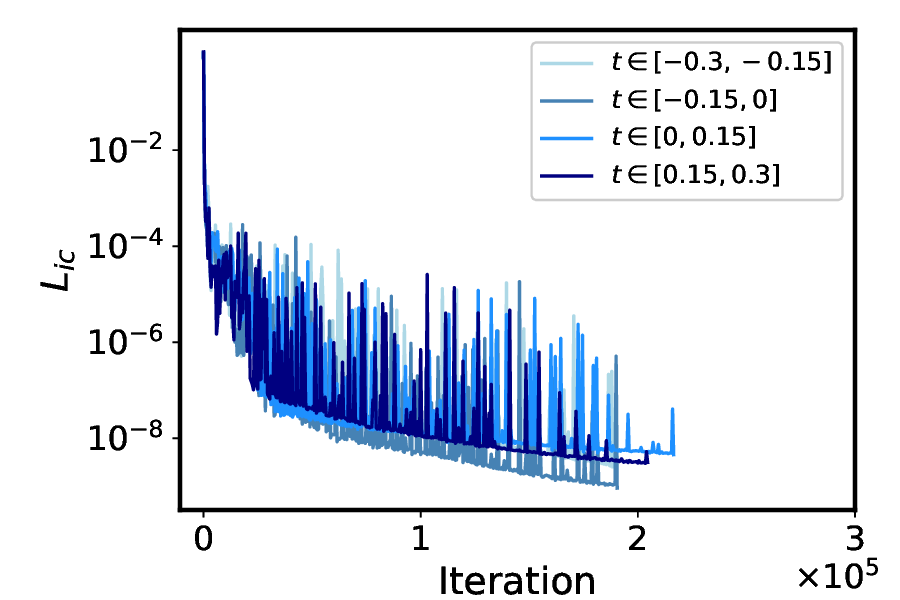}
\includegraphics[width=5.5cm,height=3.6cm]{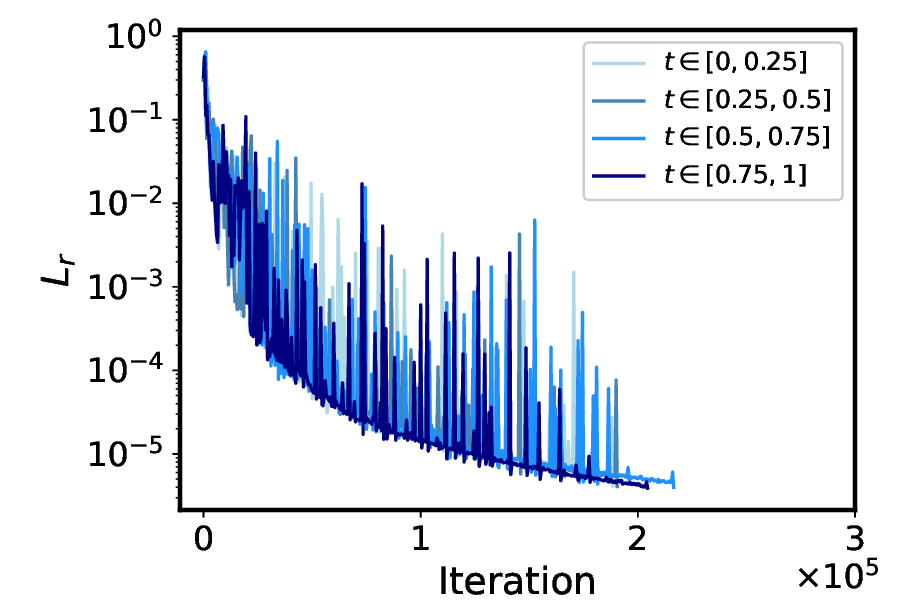}
\includegraphics[width=5.5cm,height=3.6cm]{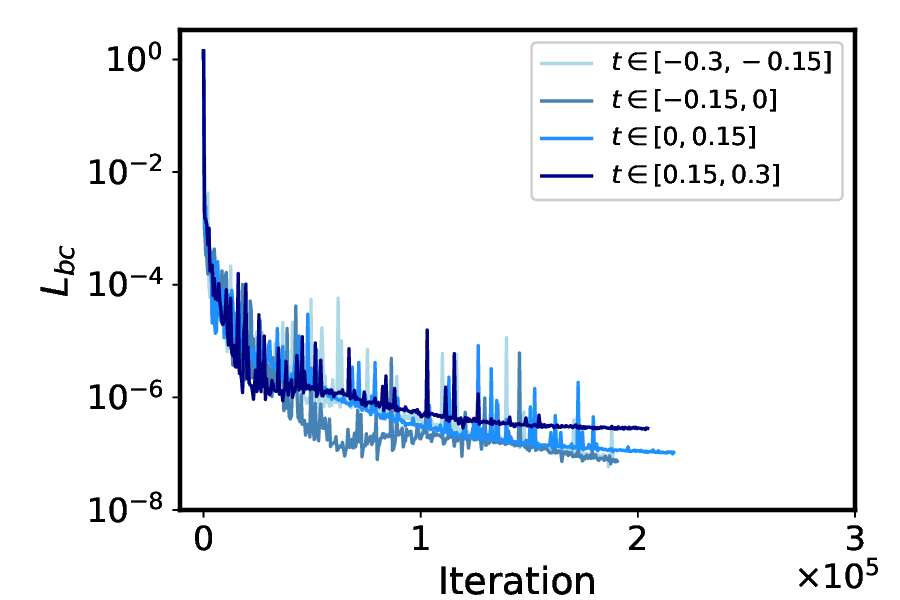}
$b$
\includegraphics[width=5.5cm,height=3.6cm]{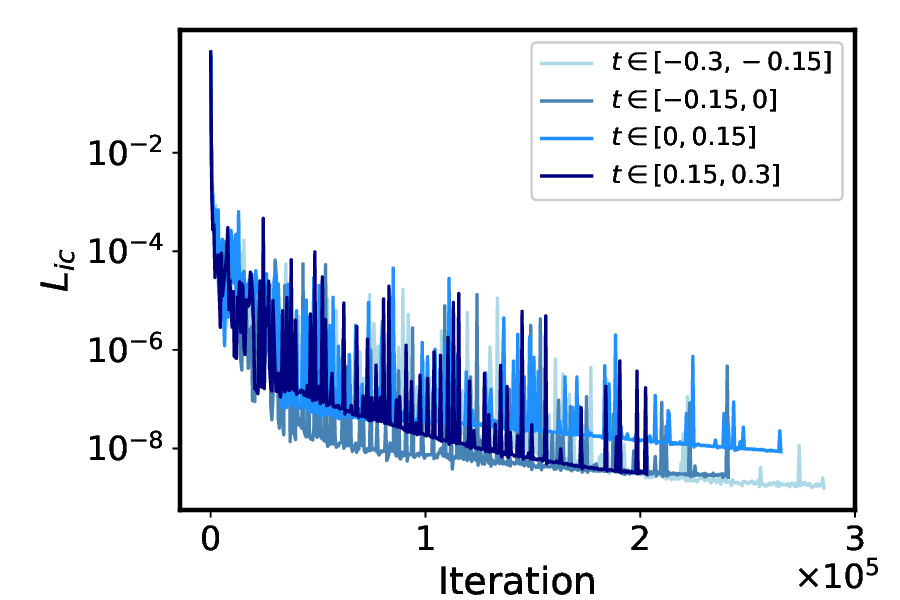}
\includegraphics[width=5.5cm,height=3.6cm]{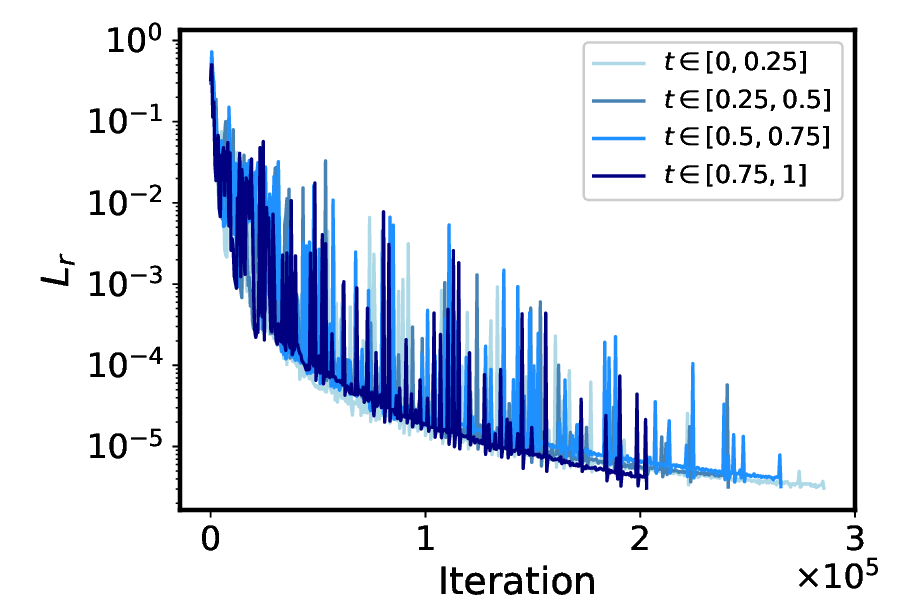}
\includegraphics[width=5.5cm,height=3.6cm]{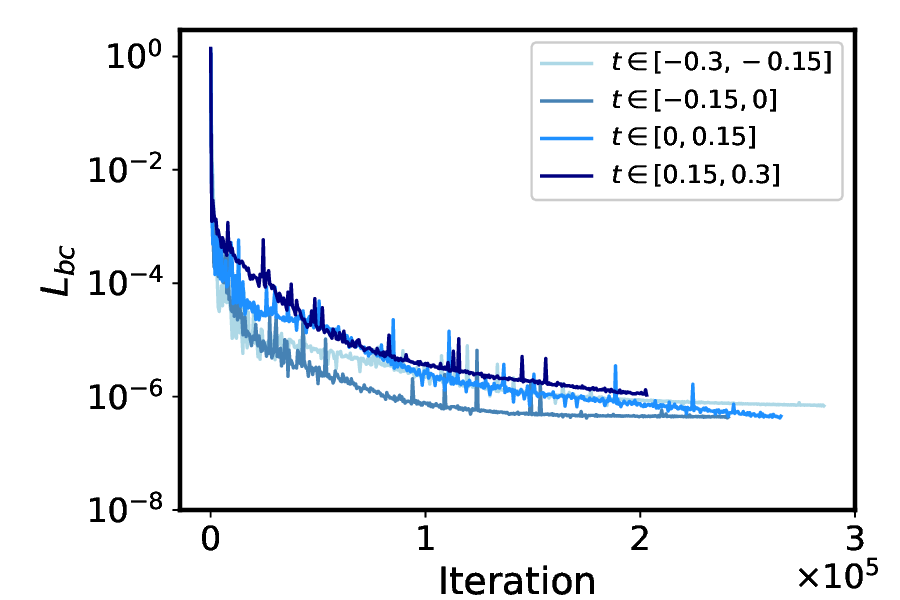}
$c$
\includegraphics[width=5.5cm,height=3.6cm]{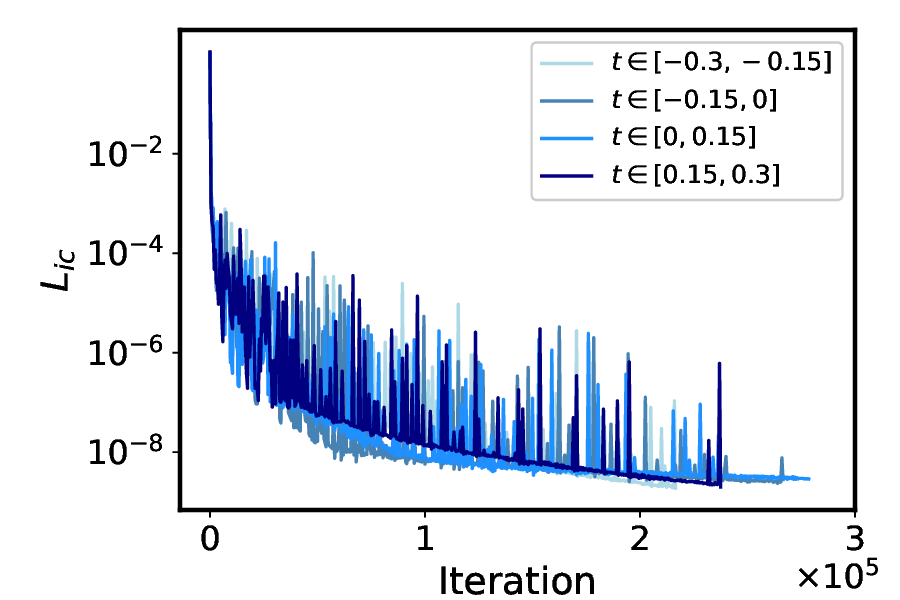}
\includegraphics[width=5.5cm,height=3.6cm]{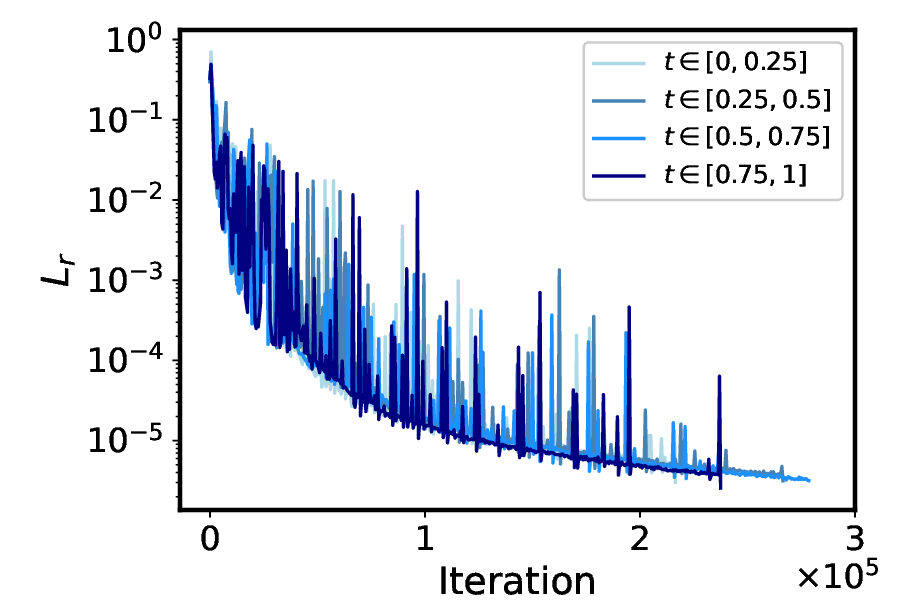}
\includegraphics[width=5.5cm,height=3.6cm]{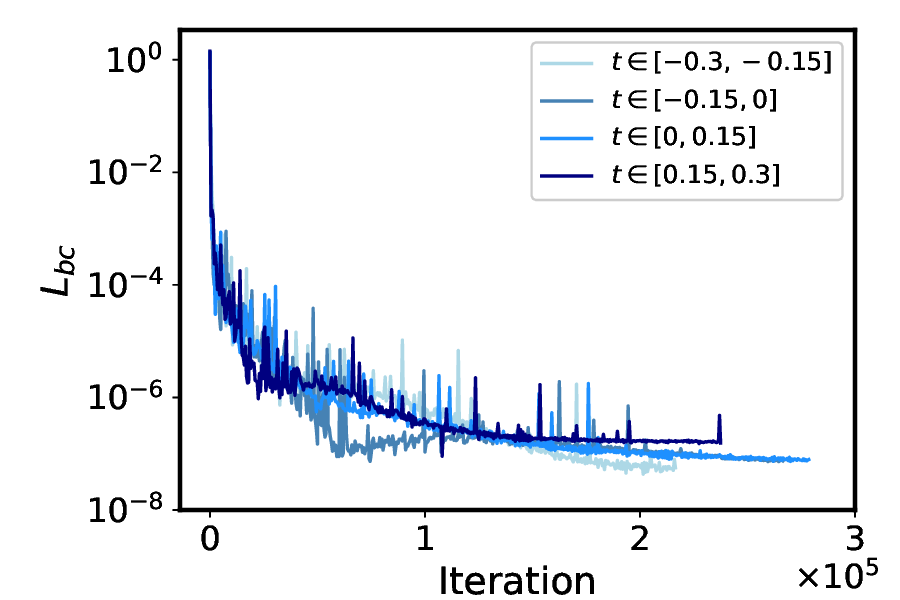}
$d$
\includegraphics[width=5.5cm,height=3.6cm]{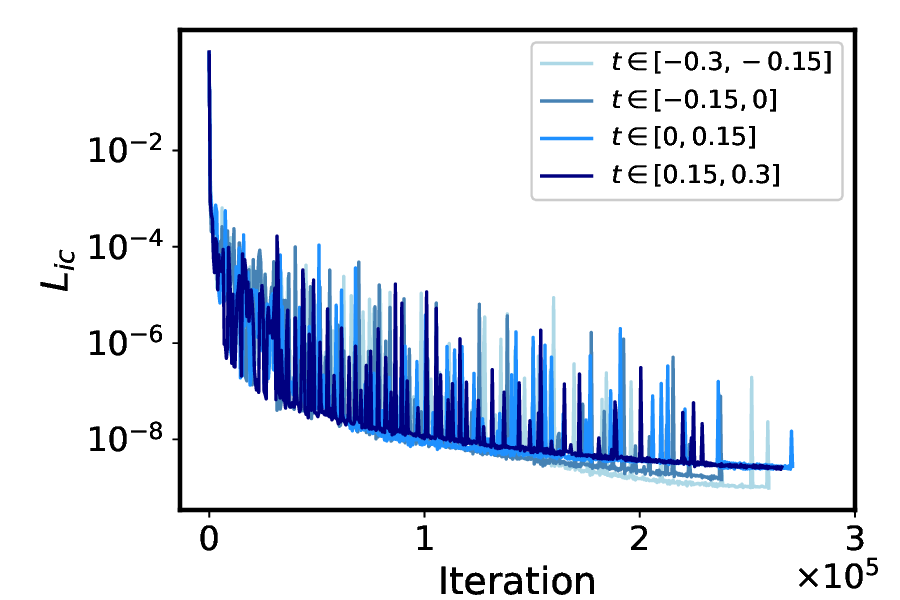}
\includegraphics[width=5.5cm,height=3.6cm]{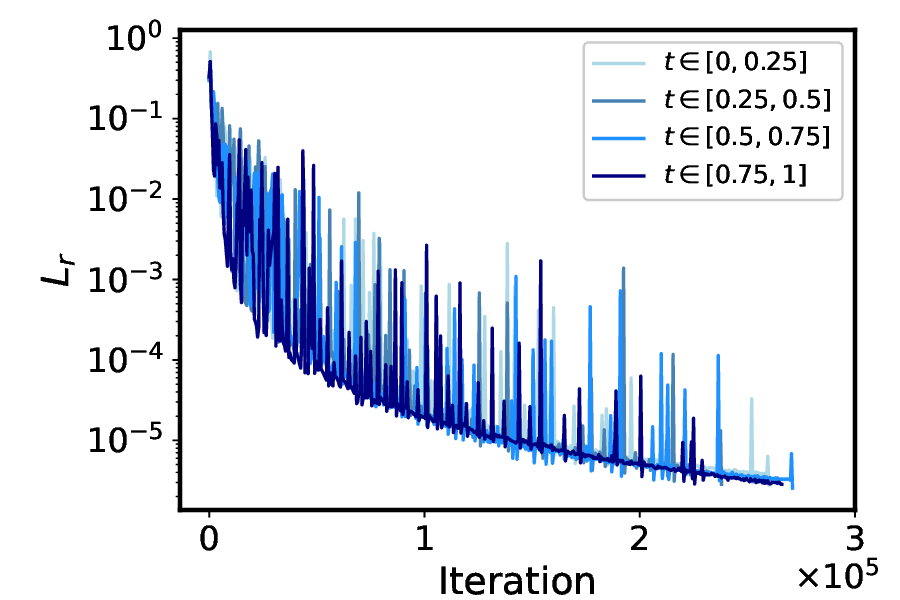}
\includegraphics[width=5.5cm,height=3.6cm]{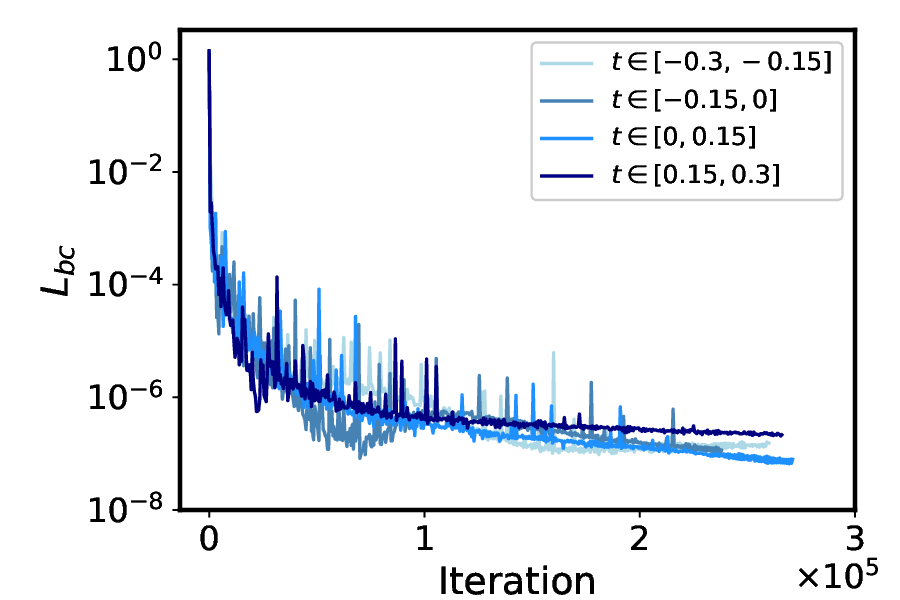}
$d$
\caption{(Color online) Loss convergence of Causal AS method with $p=1$ for mKdV equation: (a) $N_A=1000$; (b) $N_A=2000$; (c) $N_A=3000$; (d) $N_A=4000$; (e) $N_A=5000$.}
\label{figA-3}
\end{figure}

Fig. \ref{figA-3} shows the evolution curves of the loss terms across different time windows when numerically solving the mKdV equation with the Causal AS method.

\end{document}